\documentclass[aps,twocolumn,prd,showpacs,showkeys,preprintnumbers,superscriptaddress,bibnotes,floatfix]{revtex4-1}

\pdfoutput=1
\usepackage{amsmath}
\usepackage{amsfonts}
\usepackage{amssymb}
\usepackage{mathrsfs}
\usepackage{color}
\usepackage{bm}
\usepackage{graphicx}
\usepackage{blindtext}
\usepackage{wasysym}
\usepackage{mwe}
\usepackage{hyperref}
\usepackage[normalem]{ulem}
\usepackage{multirow}

\newcommand{\ie}{{\it i.e.}}

\newcommand{\eg}{{\it e.g.}}

\newcommand{\fig}{Fig.}

\newcommand{\Ref}{Ref.}
\newcommand{\Refs}{Refs.}

\newcommand{\figu}[1]{\fig~\ref{fig:#1}}

\newcommand{\bi}{\begin{itemize}}
\newcommand{\ei}{\end{itemize}}

\begin{document}
 
\title{New limits on neutrino decay\\from the Glashow resonance of high-energy cosmic neutrinos}

\author{Mauricio Bustamante}
\email{mbustamante@nbi.ku.dk}
\thanks{ORCID: \href{http://orcid.org/0000-0001-6923-0865}{0000-0001-6923-0865}}
\affiliation{Niels Bohr International Academy \& DARK, Niels Bohr Institute,\\University of Copenhagen, DK-2100 Copenhagen, Denmark}

\date{April 15, 2020}

\begin{abstract}

Discovering neutrino decay would be strong evidence of physics beyond the Standard Model.
Presently, there are only lax lower limits on the lifetime $\tau$ of neutrinos, of $\tau/m \sim 10^{-3}$~s~eV$^{-1}$ or worse, where $m$ is the unknown neutrino mass.  High-energy cosmic neutrinos, with TeV--PeV energies, offer superior sensitivity to decay due to their cosmological-scale baselines.  To tap into it, we employ a promising method, recently proposed, that uses the Glashow resonance $\bar{\nu}_e + e \to W$, triggered by $\bar{\nu}_e$ of 6.3~PeV, to test decay with only a handful of detected events.  If most of the $\nu_1$ and $\nu_2$ decay into $\nu_3$ en route to Earth, no Glashow resonance would occur in neutrino telescopes, because the remaining $\nu_3$ have only a tiny electron-flavor content.  We turn this around and use the recent first detection of a Glashow resonance candidate in IceCube to place new lower limits on the lifetimes of $\nu_1$ and $\nu_2$. 
For $\nu_2$, our limit is the current best. For $\nu_1$, our limit is close to the current best and, with the imminent detection of a second Glashow resonance, will vastly surpass it.

\end{abstract}

\maketitle


{\bf Introduction.---}  In the Standard Model (SM), neutrinos decay only with lifetimes many orders of magnitude longer than the age of the Universe\ \cite{Pal:1981rm, Hosotani:1981mq, Nieves:1982bq}.  For all practical purposes, they are stable.  Yet, in proposed SM extensions, neutrinos may decay faster by emitting new particles with which they couple strongly; see, \eg, \Refs\ \cite{Bahcall:1972my, Chikashige:1980qk, Gelmini:1982rr, Tomas:2001dh, Hannestad:2005ex, Zhou:2007zq, Chen:2007zy, Li:2007kj, Escudero:2019gfk}.  In this case, decay, though still rare, may be detectable in neutrinos that travel long distances.  Detecting it, or significantly constraining the neutrino lifetime, would help to steer SM extensions.

This makes high-energy cosmic neutrinos, with energies of TeV--PeV and traveled distances of Mpc--Gpc\ 
\cite{Aartsen:2013bka, Aartsen:2013jdh, Aartsen:2014gkd, Aartsen:2015rwa, Aartsen:2016xlq},
ideal probes of neutrino decay\ \cite{Pakvasa:1981ci, Beacom:2002vi, Barenboim:2003jm, Beacom:2003nh, Beacom:2003zg, Meloni:2006gv, Maltoni:2008jr, Bustamante:2010nq, Mehta:2011qb, Baerwald:2012kc, Pakvasa:2012db, Pagliaroli:2015rca, Bustamante:2015waa, Huang:2015flc, Shoemaker:2015qul, Bustamante:2016ciw, Rasmussen:2017ert, Ahlers:2018mkf, Denton:2018aml}.  Neutrinos emitted by astrophysical sources initially consist of a mixture of the three mass eigenstates, $\nu_1$, $\nu_2$, $\nu_3$.  If neutrinos are unstable, the heavier among them may decay into the lightest one\ \cite{Bahcall:1972my}.
During their trip to Earth, the cumulative effect of many decays
nominally grants sensitivity to lifetimes as long as $\tau \sim 10^2~{\rm s}~(m/{\rm eV})$, where $m$ is the unknown neutrino mass\ \cite{Bustamante:2016ciw}.  This is an improvement of $10^5$--$10^{13}$~s over the best current lower limits that come from solar\ \cite{Berryman:2014qha} (see also \Refs\ \cite{Joshipura:2002fb, Beacom:2002cb, Bandyopadhyay:2002xj, Picoreti:2015ika, Aharmim:2018fme}), atmospheric, and long-baseline neutrinos\ \cite{GonzalezGarcia:2008ru} (see also \Ref\ \cite{Gomes:2014yua}), and outperforms reactor\ \cite{Porto-Silva:2020gma} and accelerator\ \cite{Gago:2017zzy, Coloma:2017zpg} neutrinos.

\begin{figure}[t!]
 \centering
 \includegraphics[width=\columnwidth,height=\columnwidth]{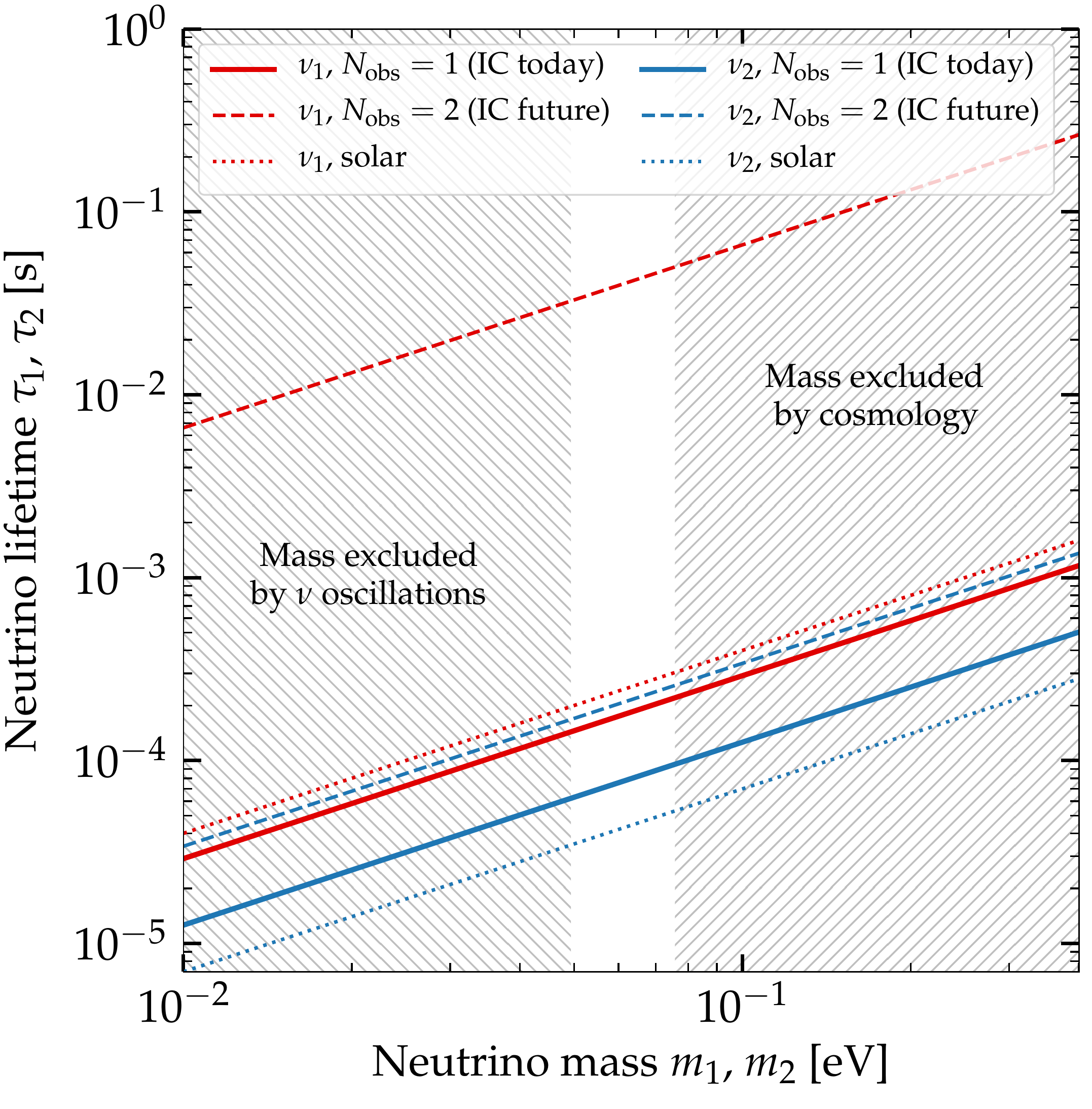}
 \caption{\label{fig:limits}
 Lower limits, at 90\%~C.L., on the lifetimes of $\nu_1$ and $\nu_2$, as a function of their masses.  We assume inverted neutrino mass ordering, \ie, $\nu_1$ and $\nu_2$ decay to a stable, visible $\nu_3$.  Our new limits come from observing $N_{\rm obs} = 1$ contained shower in the 4--8~PeV range, the first Glashow-resonance (GR) candidate, in 4.6 years of IceCube (IC)\ \cite{Talk_Lu_UHECR_2018}.  
 Projected limits come from observing 2 such showers.  Existing limits come from solar neutrinos\ \cite{Berryman:2014qha}.  Low masses are excluded by the measurement of $\Delta m_{i3}^2 \equiv m_i^2 - m_3^2$ ($i=1,2$) in oscillation experiments\ \cite{deSalas:2018bym, NuFit_4.1}; high masses, by cosmological bounds on the sum of neutrino masses\ \cite{Choudhury:2018byy} (see \Ref\ \cite{Bustamante:2016ciw} for details).}
\end{figure}

{\bf Synopsis.---}  Because the flux of high-energy cosmic neutrinos falls steeply with energy\ \cite{Aartsen:2013bka, Aartsen:2013jdh, Aartsen:2014gkd, Aartsen:2015rwa, Aartsen:2016xlq}, the rate of multi-PeV neutrinos that interact inside neutrino telescopes 
(``contained events'') via neutrino-nucleon ($\nu N$) scattering, the main detection channel, is low.  
Yet, at 6.3~PeV, $\bar{\nu}_e$ may trigger the Glashow resonance (GR), $\bar{\nu}_e + e^- \to W^-$, a long-sought SM process\ \cite{Glashow:1960zz}. 
Because at this energy the GR cross section is $\sim$200 times higher than the $\nu N$ cross section, it significantly raises the rate of  contained multi-PeV events\ \cite{Bhattacharya:2011qu, Barger:2012mz, Bhattacharya:2012fh, Barger:2014iua, Rasmussen:2017ert, Huang:2019hgs}.

Reference\ \cite{Bustamante:2016ciw} proposed using the observation of the GR to probe decay in the inverted neutrino mass ordering, where $\nu_3$ is lightest. 
If most of the cosmic multi-PeV $\nu_1$ and $\nu_2$ were to decay into $\nu_3$ en route to Earth, the remaining neutrinos would be mostly $\nu_3$.   Because $\nu_3$ has a tiny electron-flavor content, the flux would not contain sufficient $\bar{\nu}_e$ to trigger the GR within a few years in present-day detectors.
Therefore, detecting even a single event at around the GR energy would reveal the presence of $\nu_1$ or $\nu_2$, and allow us to place lower limits on their lifetimes.  
For the first time, we apply this method in a full analysis, spurred by the recent detection of the first GR candidate in the IceCube neutrino telescope\ \cite{Talk_Lu_UHECR_2018}

Figure \ref{fig:limits} shows that, for $\nu_2$, our new limit is the best one to date.  For $\nu_1$, our limit is comparable to the current best one, from solar neutrinos\ \cite{Berryman:2014qha}.  We account for particle-physics and astrophysical unknowns, and for detector effects, and ensure that our limits are conservative.


{\bf Neutrino mixing.---}  Neutrinos are created as flavor states, $\nu_e$, $\nu_\mu$, and $\nu_\tau$,  but propagate as mass eigenstates, $\nu_1$, $\nu_2$, and $\nu_3$, each with a different mass $m_i$ ($i = 1, 2, 3$), whose value is presently unknown\ \cite{Capozzi:2017ipn}.  The flavor and mass states are connected by the lepton mixing matrix $U$\ \cite{Maki:1962mu, Pontecorvo:1967fh}.  Each $\nu_i$ contains different amounts of electron, muon, and tau flavor, respectively, $\lvert U_{e i} \rvert^2$, $\lvert U_{\mu i} \rvert^2$, and $\lvert U_{\tau i} \rvert^2$.  (Unless otherwise indicated, $\nu_i$ refers to both $\nu_i$ and $\bar{\nu}_i$.)   Following convention\ \cite{Tanabashi:2018oca}, we write $U$ in terms of four mixing parameters: three angles, $\theta_{12}$, $\theta_{23}$, and $\theta_{13}$, and one CP-violation phase, $\delta_{\rm CP}$.  Their values are known experimentally\ (\eg, \Refs\ \cite{Capozzi:2018ubv, deSalas:2018bym, Esteban:2018azc}), with different precision, which we account for later.
Critical to our work is that $\lvert U_{e3} \rvert^2 \lesssim 5\%$; the tiny electron-flavor content of $\nu_3$ make it unlikely to trigger a GR.

While propagating, neutrinos oscillate: a neutrino created as $\nu_\alpha$ may be detected later as $\nu_\beta$ ($\alpha, \beta = e, \mu, \tau$).  Formally, the probability $P_{\alpha\beta}$ that this occurs depends on the distance $L$ traveled by the neutrino and on its energy $E_\nu$.  However, for high-energy cosmic neutrinos, because oscillations are rapid, we are sensitive only to the average probability, $P_{\alpha \beta} = \sum_i \lvert U_{\alpha i} \rvert^2 \lvert U_{\beta i} \rvert^2$\ \cite{Pakvasa:2008nx}.


{\bf Flavor ratios.---}  In astrophysical sources, high-energy protons interact with ambient matter\ \cite{Margolis:1977wt, Stecker:1978ah, Kelner:2006tc} and photons\ \cite{Stecker:1978ah, Mucke:1999yb, Hummer:2010vx} to produce pions.  Their decay ($\pi^+ \to \mu^+ + \nu_\mu$) and the subsequent decay of muons ($\mu^+ \to \bar{\nu}_\mu + e^+ + \nu_e$) produce high-energy neutrinos.  This yields the nominal expectation for the flavor ratios, $(f_e:f_\mu:f_\tau)_{\rm{S}} = \left( \frac{1}{3}:\frac{2}{3}:0 \right)$, where $f_\alpha$ is the ratio of $\nu_\alpha + \bar{\nu}_\alpha$ to the total.   Oscillations change the flavor ratios into $f_{\alpha,\oplus} = \sum_\beta P_{\beta\alpha} f_{\beta,\text{S}}$ upon reaching Earth.  For the nominal expectation,
this yields $(f_e:f_\mu:f_\tau)_\oplus \approx \left( \frac{1}{3}:\frac{1}{3}:\frac{1}{3} \right)$.
However, there are large uncertainties in the production\ \cite{Bustamante:2019sdb}, so below we explore all possible combinations of $f_{e, {\rm S}}$, $f_{\mu, {\rm S}}$, and $f_{\tau, {\rm S}}$. 
Since we focus on a narrow energy range (4--8~PeV), we assume that $f_{\alpha, {\rm S}}$ are constant.


{\bf Neutrino decay.---}  We adopt a generic scenario of non-radiative (\ie, without photons) neutrino decay where the daughter neutrino is visible, \ie, detectable, in neutrino telescopes.  
Following the method outlined above, we assume the inverted neutrino mass ordering, where $\nu_3$ is lightest; we take it to be stable.
We let $\nu_1$ and $\nu_2$ decay via $\nu_1 \to \nu_3 + \phi$ and $\nu_2 \to \nu_3 + \phi$, where $\phi$ is a light new boson, \eg, a Majoron\ \cite{Schechter:1981cv, Gelmini:1982rr, Tomas:2001dh}, without definite lepton number, or a lepton number-carrying scalar\ \cite{Berryman:2018ogk}.  
The Majorana or Dirac nature of neutrinos determines what helicities are available to the daughter neutrino, which in turn determines whether it is visible\ \cite{deGouvea:2019goq}. 
We focus on the likely case of Majorana neutrinos, which entails no helicity suppression for the daughters.  

We assume that the daughter $\nu_3$ receives the full parent energy\ \cite{Bustamante:2016ciw}.
Hence, decay merely converts multi-PeV $\nu_1$ and $\nu_2$ into multi-PeV $\nu_3$, which may still trigger the GR, albeit at a very low rate.  
Under the alternative assumption that $\nu_3$ receives a fraction of the parent energy, then, because the neutrino flux falls steeply with energy, no multi-PeV $\nu_3$ would be left to trigger the GR.
Thus, the former assumption is more compatible with the observation of multi-PeV events from which we derive lifetime limits.  We adopt it to ensure our limits are conservative.

Other than the above assumptions, our limits on neutrino lifetimes are model-independent.  Later, we translate them into limits on the interaction with $\phi$ assuming that it has scalar and pseudoscalar couplings\ \cite{Bahcall:1972my, Beacom:2002cb}, described by $\mathcal{L} = g_{ij} \bar{\nu}_i \nu_j \phi + h_{ij} \bar{\nu}_i \gamma_5 \nu_j \phi + {\rm h.c.}$, where $g_{ij}$ and $h_{ij}$ are coupling constants.  More sophisticated models exist\ \cite{deGouvea:2019goq}, but we do not explore them here.

The inverted mass ordering remains viable though there are indications that it is normal\ \cite{Capozzi:2020qhw}.  In the normal ordering, the decay of $\nu_2$ and $\nu_3$ into $\nu_1$ is better probed with high-energy cosmic neutrinos via flavor ratios\ \cite{Bustamante:2016ciw}.


{\bf Decay in high-energy cosmic neutrinos.--}  After traveling a distance $L$, the number of remaining unstable $\nu_i$ of energy $E_\nu$, with lifetime $\tau_i$, is reduced by a factor of $\exp[-(m_i/\tau_i)(L/E_\nu)]$.  
Here, $E_\nu/m_i$ is the Lorentz boost of the neutrino; in the lab frame, the neutrino lifetime is $(E_\nu/m_i) \tau_i$, so more energetic neutrinos live longer.
Thus, PeV-scale cosmic neutrinos with known $L$ and $E_\nu$ are nominally sensitive to $\tau_i/m_i \sim 10^2~(L/{\rm Gpc}) ({\rm PeV}/E_\nu)$~s~eV$^{-1}$. 
Since $\tau_i$ appears in the ratio $\tau_i/m_i$, below we place limits on this ratio.

The cosmological expansion dampens the energy of neutrinos emitted by a distant source located at redshift $z$, which affects their lifetime in the lab frame.  We follow \Refs\ \cite{Baerwald:2012kc, Bustamante:2016ciw} to incorporate the effects of decay, including redshift corrections, into the flavor ratios, \ie, $f_{\alpha,\oplus} \equiv f_{\alpha,\oplus} \left( E_\nu, z; f_{e, {\rm S}}, f_{\mu, {\rm S}}, \boldsymbol\theta, \tau_1/m_1, \tau_2/m_2 \right)$, where $\boldsymbol\theta \equiv (\theta_{12}, \theta_{23}, \theta_{13}, \delta_{\rm CP})$.  The Supplemental Material contains the full expressions.


{\bf Testing decay via Glashow.---}  In the case where most of the $\nu_1$ and $\nu_2$ have decayed before reaching Earth, the flux contains mostly $\nu_3$, and the flavor ratios are given by its flavor content, \ie, $f_{\alpha, \oplus} \approx \lvert U_{\alpha 3} \rvert^2$.  
Because $f_{e, \oplus} \approx \lvert U_{e3} \rvert^2 \ll 1$, the number of $\bar{\nu}_e$ arriving at Earth would be too low to yield a detectable rate of GR events at neutrino telescopes within a few years.  Therefore, detecting even a single GR event would allow us to place lower limits on the lifetimes of $\nu_1$ and $\nu_2$.
Reference\ \cite{Bustamante:2016ciw} showed the promise of this method, but did so assuming that decay was complete upon reaching Earth, using only two representative flux cases, and a simplified computation of event rates.  Below, we lift these simplifications.


{\bf Diffuse neutrino flux.---}  The astrophysical sources responsible for the bulk of the observed diffuse flux of high-energy neutrinos are unknown, but likely extragalactic\ \cite{Ahlers:2013xia, Ahlers:2015moa, Murase:2015xka, Denton:2017csz, Aartsen:2017ujz, Ahlers:2018fkn}.  We compute the flux, including the effects of neutrino decay, as coming from a population of unspecified extragalactic sources whose number density $\rho_{\rm src}$ evolves with redshift.  We add the contributions of sources up to $z=4$;  more distant ones contribute negligibly.  The contribution of each source to the energy flux of $\nu_\alpha$ at Earth is $J_{\nu_\alpha} \equiv \phi_0 (1-f_{\bar{\nu}}) f_{\alpha, \oplus} E_\nu^{2-\gamma}$, where the normalization $\phi_0$, the spectral index $\gamma$, and the fraction $f_{\bar{\nu}}$ of $\bar{\nu}$ in the flux are free parameters that we vary below.  For $\bar{\nu}_\alpha$, we replace $(1-f_{\bar{\nu}})$ by $f_{\bar{\nu}}$.
The neutrino luminosity density $\rho_{\rm src} J_{\nu_\alpha}$ follows the star formation rate\ \cite{Yuksel:2008cu}: most candidate sources lie  at $z \approx 1$, or 2--3~Gpc\ \cite{Anchordoqui:2013dnh}.
The Supplemental Material has details of the calculation.

At low energies, more $\nu_1$ and $\nu_2$ decay into $\nu_3$, and so the flux of $\bar{\nu}_e$ is lower.  At high energies, lifetimes are longer and the $\bar{\nu}_e$ flux is higher.
Our analysis is sensitive to lifetimes shorter than $10^3$~s~eV$^{-1}$, for which decay is complete or significant at the GR energy of 6.3~PeV.


{\bf In-Earth propagation.---}  Once neutrinos reach Earth, we propagate them along all directions through its interior, where they interact with matter, and up to IceCube, located at the South Pole.  
Neutral-current (NC) $\nu N$ deep inelastic scatterings ($\nu_\alpha + N \to \nu_\alpha + X$, where $X$ are hadrons) dampen the flux at high energies, since final-state neutrinos escape undetected with 70\% of the parent neutrino energy, on average.  Charged-current (CC) scatterings ($\nu_\alpha + N \to \alpha + X$) attenuate the flux by removing neutrinos.   (The CC scattering of a $\nu_\tau$ produces a tauon that decays into a $\nu_\tau$, so the $\nu_\tau$ flux is less attenuated.)
The GR attenuates the $\bar{\nu}_e$ flux around 6.3~PeV.

We use {\tt nuSQuIDS}\ \cite{Delgado:2014kpa, SQuIDS, NuSQuIDS} to propagate neutrinos inside Earth along each direction $\cos \theta_z$, where $\theta_z$ is the zenith angle measured from the South Pole.  For the matter density profile, we use the Preliminary Reference Earth Model\ \cite{Dziewonski:1981xy}.  The effect of in-Earth propagation on the neutrino spectrum varies for each flavor, for $\nu$ and $\bar{\nu}$, and is more significant for high energies and long paths.  It is important only for upgoing neutrinos, \ie, $\cos \theta_z < 0$.


\begin{figure}[t!]
 \centering
 \includegraphics[width=\columnwidth,height=\columnwidth]{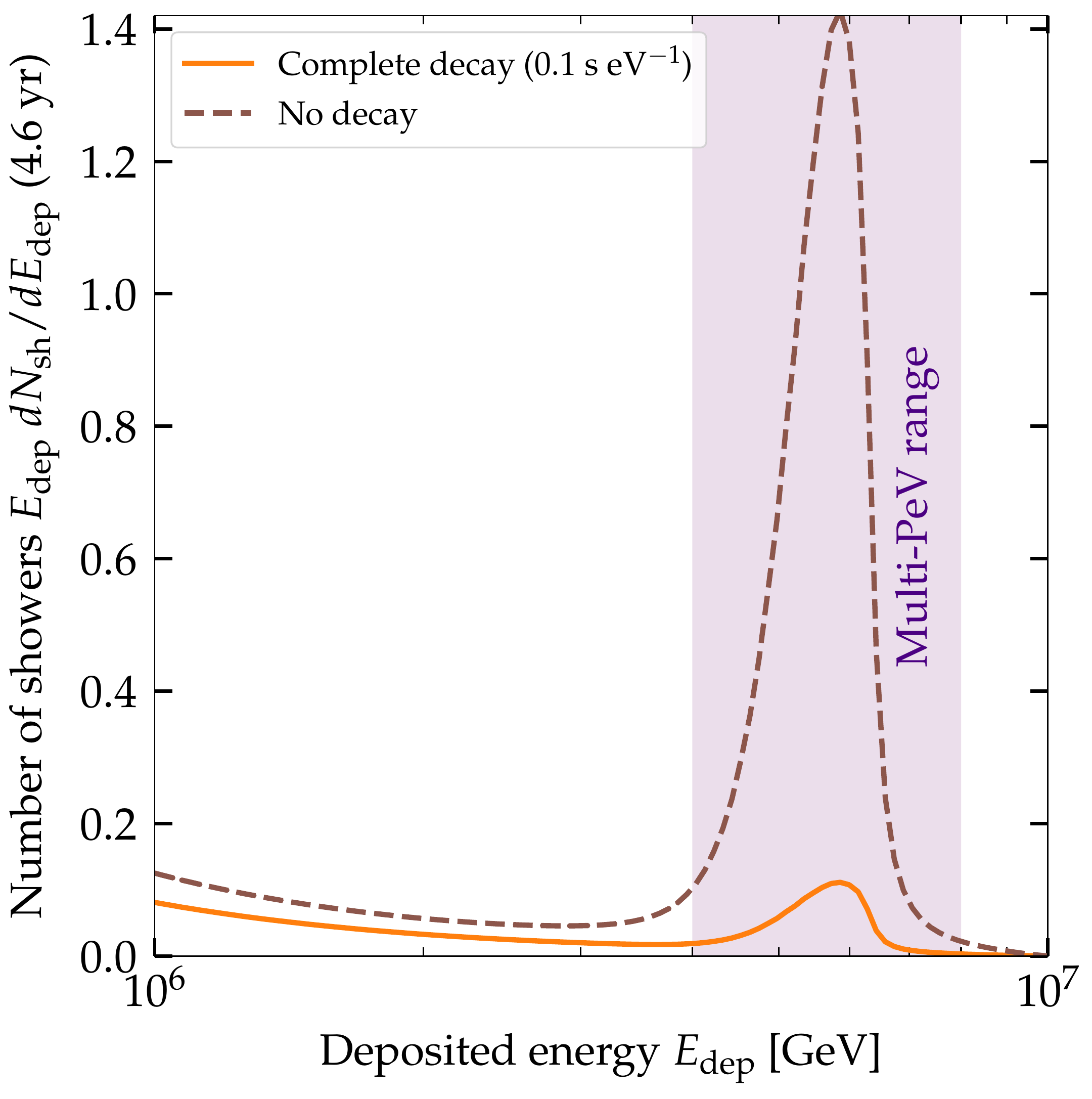}
 \caption{\label{fig:shower_spectrum}
 Number of contained multi-PeV showers in IceCube in 4.6 years of exposure, without decay and with complete decay.  For this plot, we choose illustrative values of $\gamma = 2.1$, $\phi_0 = 8 \cdot 10^{-7}$~GeV~cm$^{-2}$~s$^{-1}$~sr$^{-1}$, $(f_e:f_\mu:f_\tau)_{\rm S} = (\frac{1}{3}:\frac{1}{3}:\frac{1}{3})$, $f_{\bar{\nu}} = 0.5$, best-fit values of the mixing parameters, and equal lifetimes for $\nu_1$ and $\nu_2$, of $\tau_1/m_1 = \tau_2/m_2 = 0.1$~s~eV$^{-1}$.  In our analysis, we let these parameters vary; see the main text.}
\end{figure}

{\bf Detecting high-energy neutrinos.---}  Presently, IceCube is the largest neutrino telescope, an array of photomultipliers that instruments 1~km$^3$ of Antarctic ice at depths of 1.5--2.5~km.  When a high-energy neutrino scatters off the ice, it creates final-state charged particles that radiate Cherenkov light.  From the amount of light collected, it is possible to infer the neutrino energy.

For a given neutrino flux at IceCube, we compute the expected detection rate via $\nu N$ NC and CC scattering of all flavors of $\nu$ and $\bar{\nu}$, and via the GR of $\bar{\nu}_e$. 
We focus on ``shower'' events born from all of these interactions, \ie, particle showers around the interaction point, with a light profile that expands roughly spherically.
This is the most likely outcome of the decay of the $W$ from a GR: 67\% of the time it decays into hadrons, 11\% into electrons, and 11\% into tauons, all of which shower promptly\ \cite{Tanabashi:2018oca}.  

The first GR candidate was found by IceCube in 4.6~years of data, in the form of a partially contained shower with an energy of about 6~PeV\ \cite{Talk_Lu_UHECR_2018}.  Preliminary results show that the most likely energy of the parent neutrino matches the GR energy\ \cite{Talk_Lu_UHECR_2018}.  Its identity as a GR shower, and not a $\nu N$ CC shower, is further supported by a new analysis that hints at the presence of muons in numbers that are compatible with an origin in the hadronic decay of the $W$ made in a GR\ \cite{Talk_Lu_UHECR_2018, Lu:ICRC2019}.


{\bf Multi-PeV shower rate.---}  The energy $E_{\rm dep}$ deposited by a shower 
depends on the energy $E_\nu$ of the parent neutrino and on what fraction of it is given to the final-state charged particles, \ie, on the inelasticity.
For a given neutrino flux, we follow the procedure from \Ref\ \cite{Palomares-Ruiz:2015mka} to compute the shower spectrum $dN_{\rm sh}/dE_{\rm dep}$ due to the GR plus $\nu N$ interactions.  It accounts for the different relation between $E_\nu$ and $E_{\rm dep}$ for each interaction type and decay channel of final-state tauons and $W$ bosons, and for the IceCube energy resolution, of $\sim$12\%\ \cite{Aartsen:2013vja}.  The Supplemental Material outlines the calculation.

Our observable is the expected number $N_{\rm sh}$ of showers with $E_{\rm dep} = 4$--8~PeV, detected over a time $T$, and integrated over all arrival directions, \ie,
\begin{equation*}
 N_{\rm sh}
 =
 2\pi T
 \int_{4~{\rm PeV}}^{8~{\rm PeV}} dE_{\rm dep}
 \int_{-1}^1 d\cos\theta_z
 \frac{dN_{\rm sh}(E_{\rm dep}, \cos\theta_z)}{dE_{\rm dep}} \;,
\end{equation*}
for given values of the free parameters of our analysis.
At these energies, the contribution of atmospheric neutrinos is unimportant\ \cite{Beacom:2004jb}, so we neglect it.

Figure\ \ref{fig:shower_spectrum} shows the shower spectrum, without decay and with complete decay, for an illustrative choice of parameters.  In it, under complete decay, the multi-PeV shower rate is reduced by a factor of $\left\vert U_{e3} \right\vert^2/f_{e,\oplus}$.  For the nominal expectation of $f_{e,\oplus} \approx 1/3$, this factor is about 0.1.  In this case, the probability that a fluctuation yields one or more multi-PeV showers is a few tens of percent\ \cite{Bustamante:2016ciw}.
Below, we explore a wide variety of values of $\left\vert U_{e3} \right\vert^2$ and $f_{e,\oplus}$, which significantly alter this probability.


{\bf Statistical analysis.---}  We generate test shower rates $N_{\rm sh}$ for many different choices of values of the free parameters, which are listed below, and compare them to the number of contained showers $N_{\rm obs}$ observed by IceCube in the same range of 4--8~PeV.  When computing present-day limits on the lifetimes, we set $N_{\rm obs} = 1$ in $T = 4.6$~years\ \cite{Talk_Lu_UHECR_2018}.  When making projections, we scale up these numbers assuming that they reflect the rate of multi-PeV showers, \ie, one in IceCube every 4.6~years.

We compare the expected and observed shower rates via the unbinned Poissonian likelihood function
\begin{equation*}
 \vspace*{-0.01cm}
 \mathcal{L}\left( \phi_0, \gamma, f_{e, {\rm S}}, f_{\mu, {\rm S}}, f_{\bar{\nu}}, \boldsymbol\theta, \frac{\tau_1}{m_1}, \frac{\tau_2}{m_2}; N_{\rm obs}\right)
 =
 \frac{e^{-N_{\rm sh}} 
 N_{\rm sh}^{N_{\rm obs}}} {N_{\rm obs}!} \;,
\end{equation*}
where $N_{\rm sh} \equiv N_{\rm sh}( \phi_0, \gamma, f_{e, {\rm S}}, f_{\mu, {\rm S}}, f_{\bar{\nu}}, \boldsymbol\theta, \tau_1/m_1, \tau_2/m_2 )$ depends on the 11 free parameters of the analysis.  (Since $f_{\tau, {\rm S}} = 1-f_{e, {\rm S}}-f_{\mu, {\rm S}}$, we need only vary $f_{e, {\rm S}}$ and $f_{\mu, {\rm S}}$.)  We vary their values independently of each other.  The likelihood accounts for the possibility that in a flux depleted of $\bar{\nu}_e$ by decay, upward random fluctuations in the $\nu N$ rate mimic true GR showers.  We adopt a Bayesian approach to maximize the likelihood and use {\tt MultiNest}\ \cite{Feroz:2007kg, Feroz:2008xx, Feroz:2013hea, Buchner:2014nha} to efficiently explore the parameter space.  

For the flux normalization and spectral index, $\phi_0$ and $\gamma$, we use priors based on the most recent measurement of the IceCube diffuse flux at lower energies using $\nu_\mu$\ \cite{Haack:2017dxi}, which, extrapolated, is compatible with the GR candidate\ \cite{Talk_Lu_UHECR_2018}.
For the mixing parameters, $\boldsymbol\theta$, we use priors built from the recent {\tt NuFit}~4.1 global fit to oscillation data, assuming inverted mass ordering\ \cite{deSalas:2018bym, NuFit_4.1}.  
For $f_{e,{\rm S}}$, $f_{\mu, {\rm S}}$, and $f_{\bar{\nu}}$, we use uniform priors across their full ranges of values.
For the lifetimes, we use generous uniform priors in $\log_{10}[(\tau_j/m_j)/({\rm s}~{\rm eV}^{-1})] \in [-5,15]$ ($j=1,2$).  The Supplemental Material contains details.


{\bf Results.---}  Presently, with $N_{\rm obs} = 1$ observed shower, we find no statistically significant evidence for neutrino decay.  Already at 90\%~C.L., the lifetimes are only bounded from below.  (The Bayes factor comparing the Bayesian evidence of our fit to a fit without neutrino decay is $B \approx 0.27$, which, in Jeffreys' scale\ \cite{Jeffreys:1939xee}, means that the scenario without decay is favored.)  Therefore, after marginalizing over all other parameters, we set the following lower limits on the lifetimes:
\begin{eqnarray*}
 \tau_1/m_1 &>& 2.91 \cdot 10^{-3} ~{\rm s}~{\rm eV}^{-1} ~(90\%~{\rm C.L.})\;, \\
 \tau_2/m_2 &>& 1.26 \cdot 10^{-3} ~{\rm s}~{\rm eV}^{-1} ~(90\%~{\rm C.L.})\;.
\end{eqnarray*}
The limit for $\nu_1$ is better because its electron-flavor content is larger than that of $\nu_2$, so we are more sensitive to the decay of $\nu_1$.
Compared to the estimated sensitivity of 10~s~eV$^{-1}$ from \Ref\ \cite{Bustamante:2016ciw}, these limits are lower because they account for parameter uncertainties, some of which are large.  Below we show that higher statistics quickly match and surpass the estimate from \Ref\ \cite{Bustamante:2016ciw} for $\nu_1$.

Figure\ \ref{fig:limits} shows that our limit for $\nu_1$ is comparable to, but slightly worse, than the limit of $\tau_1/m_1 > 4 \cdot 10^{-3}$~s~eV$^{-1}$, from the invisible decay of solar neutrinos\ \cite{Berryman:2014qha}, while our limit for $\nu_2$ is the best to date, 80\% better than the limit of $\tau_2/m_2 > 7 \cdot 10^{-4}$~s~eV$^{-1}$\ \cite{Berryman:2014qha}.  

\begin{table}[t!]
 \begin{ruledtabular}
  \caption{\label{tab:fit_parameters_future}Present-day and projected lower limits on the lifetimes of $\nu_1$ and $\nu_2$, based on the observation of $N_{\rm obs}$ showers with 4--8~PeV in IceCube.  For each $\tau_i/m_i$, its allowed range is marginalized over all other parameters.  Our main result is highlighted.  In two cases only, we fixed $f_{\alpha, {\rm S}}$ and $f_{\bar{\nu}}$.  
  For the combined IceCube + IceCube-Gen2 projections, we set the IceCube-Gen2 volume to 5 times that of IceCube and fix the mixing parameters to their current best-fit values from {\tt NuFit}~4.1\ \cite{deSalas:2018bym, NuFit_4.1}, assuming inverted neutrino mass ordering, and including Super-Kamiokande atmospheric data.}
  \centering
  \begin{tabular}{ccccc}
   \multirow{2}{*}{$N_{\rm obs}$} & \multirow{2}{*}{$(f_e:f_\mu:f_\tau)_{\rm S}$} & \multirow{2}{*}{$f_{\bar{\nu}}$} & \multicolumn{2}{c}{Lower limit (90\%~C.L.)$~\left[\frac{\rm s}{\rm eV}\right]$} \\
                 &                               &                 & $\tau_1/m_1$ & $\tau_2/m_2$ \\
   \hline\hline
   \multicolumn{5}{c}{Present-day IceCube (4.6~years)} \\
   \hline
   {\bf 1}             & {\bf Free}                      & {\bf Free}        & \boldmath{$2.91 \cdot 10^{-3}$}            & \boldmath{$1.26 \cdot 10^{-3}$} \\               
   1             & $(\frac{1}{3}:\frac{2}{3}:0)$ & 0.5             & $1.00 \cdot 10^{-3}$                       & $5.35 \cdot 10^{-3}$ \\
   1             & $(0:1:0)$                     & 0.5             & $1.89 \cdot 10^{-4}$                       & $4.40 \cdot 10^{-2}$ \\
   \hline\hline
   \multicolumn{5}{c}{Projections IceCube (9.2, 13.8, 18.4~years)} \\
   \hline
   2             & Free                      & Free        & 0.66                                       & $3.40 \cdot 10^{-3}$ \\
   3             & Free                      & Free        & 93.92                                      & $4.57 \cdot 10^{-3}$ \\               
   4             & Free                      & Free        & 593.83                                     & $5.83 \cdot 10^{-3}$ \\               
   \hline\hline
   \multicolumn{5}{c}{Projections IceCube (18.4~years) + IceCube-Gen2 (2 years)} \\
   \hline
   6             & Free                      & Free        & $6.29 \cdot 10^3$                          & $1.20 \cdot 10^{-3}$ \\               
   \end{tabular}
 \end{ruledtabular}
\end{table}

If there is hierarchy of masses, with $m_1, m_2 \gg m_3$, the decay rate of $\nu_j$ is\ \cite{Beacom:2002cb} $\Gamma_j \equiv \tau_j^{-1} = (g_{j3}^2 + h_{j3}^2) m_j / (32 \pi)$, where $g_{j3}$ and $h_{j3}$ are the scalar and pseudoscalar couplings.  Our lifetime limits imply upper limits on the combined couplings $(g_{j3}^2 + h_{j3}^2)^{1/2}$ of $4.77 \cdot 10^{-6} ({\rm eV}/m_1)$ for $\nu_1$ and $7.24 \cdot 10^{-6} ({\rm eV}/m_2)$ for $\nu_2$, at 90\%~C.L.  

Table\ \ref{tab:fit_parameters_future} shows how fixing production properties to theory expectations affects the limits.
First, at multi-PeV energies, comparable numbers of $\nu$ and $\bar{\nu}$ may be produced (see, \eg, \Ref\ \cite{Hummer:2010vx}); we fix $f_{\bar{\nu}} = 0.5$ for testing.
Second, the electron fraction $f_{e,\oplus}$ at Earth is higher for the nominal expectation of $\left( \frac{1}{3}:\frac{2}{3}:0 \right)_{\rm S}$, coming from the full pion decay chain, than for the alternative benchmark $(0:1:0)_{\rm S}$, coming from a scenario with energy-dampened intermediate muons; see, \eg, \Refs\ \cite{Bustamante:2015waa, Bustamante:2019sdb}.  
As a result, in the muon-damped case, fewer $\nu_1$ arrive at Earth, so showers are more likely due to $\nu_2$ than to $\nu_1$.
Accordingly, Table\ \ref{tab:fit_parameters_future} shows that by fixing the flavor ratios to the muon-damped case, our analysis becomes more sensitive to the $\nu_2$ lifetime than to the $\nu_1$ lifetime.

As a by-product, we extract the fraction of $\bar{\nu}$ in the multi-PeV cosmic neutrino flux: $f_{\bar{\nu}} = 0.64 \pm 0.23$.  However, the evidence is weak, and almost the full range of $f_{\bar{\nu}} \in [0,1]$ is allowed already at $2\sigma$.  Reference\ \cite{Talk_Lu_UHECR_2018} has complementary preliminary results on the $\nu_e/\bar{\nu}_e$ ratio.


{\bf Outlook.---}  Table\ \ref{tab:fit_parameters_future} shows that the imminent observation of one additional multi-PeV shower in IceCube, \ie, a total of $N_{\rm obs} = 2$, will place the strongest limit to date on the $\nu_1$ lifetime (see also \Ref\ \cite{Huang:2018nxj}).
Observing $N_{\rm obs} = 3$ showers will match the sensitivity originally estimated in \Ref\ \cite{Bustamante:2016ciw}.  Observing $N_{\rm obs} = 4$ showers will realize the potential of high-energy cosmic neutrinos to test decay, matching the ``ultimate IceCube sensitivity''\ \cite{Bustamante:2016ciw}, but without relying on knowing the precise distance to the neutrino sources.
Our limits fall shy only of the sensitivity of $10^5$~s~eV$^{-1}$ or better from core-collapse supernovae\ \cite{Ando:2003ie, Fogli:2004gy, deGouvea:2019goq} that, however, is hampered by uncertainties in neutrino emission and mixing.  
(If $\phi$ is massless or very light, there are strong limits from cosmology\ \cite{Beacom:2004yd, Hannestad:2005ex, Serpico:2007pt,  Escudero:2019gfk} and astrophysics\ \cite{Kachelriess:2000qc, Farzan:2002wx}, but this is discouraged by recent neutrino mass limits\ \cite{Beacom:2004yd, Bustamante:2016ciw}.)

IceCube-Gen2\ \cite{Aartsen:2019swn}, an upgrade planned for the 2030s, will have a volume 5--7 times larger. It should detect about one multi-PeV shower per year.  By then, the mixing parameters should be known precisely; in our projections, we fix them to their present best-fit values\ \cite{NuFit_4.1}.  Assuming that IceCube detects 4 showers in 18.4~years (2011--2030) and IceCube-Gen2 detects 2 more in 2 years (2030--2032), the $\nu_1$ limit would be $\tau_1/m_1 \gtrsim 6 \cdot 10^3$~s~eV$^{-1}$, nearly six orders of magnitude over our present limit.


{\bf Summary.---}  We have placed new limits on the lifetimes of the neutrinos $\nu_1$ and $\nu_2$ using a novel method: the observation of the first Glashow-resonance candidate in IceCube at multi-PeV energies.  We assumed an inverted neutrino mass hierarchy in which $\nu_1$ and $\nu_2$ decay into a visible $\nu_3$.  We factored in particle-physics, astrophysical, and detector uncertainties.  For $\nu_2$, our limit is the best to date.  For $\nu_1$, we already match the level of the current best limit.  In the near future, with just one more event detected, we will greatly surpass it.


\smallskip
{\bf Acknowledgements.}  We thank John Beacom, Andr\'e de Gouv\^{e}a, Shirley Li, Kohta Murase, and Anna Suliga for a careful reading of the manuscript and useful suggestions.  We thank the IceCube Collaboration for their work on the Glashow resonance.  MB is supported by the \textsc{Villum Fonden} project no.~13164.  This work used resources provided by the High Performance Computing Center at the University of Copenhagen.



\begin{thebibliography}{109}%
\makeatletter
\providecommand \@ifxundefined [1]{%
 \@ifx{#1\undefined}
}%
\providecommand \@ifnum [1]{%
 \ifnum #1\expandafter \@firstoftwo
 \else \expandafter \@secondoftwo
 \fi
}%
\providecommand \@ifx [1]{%
 \ifx #1\expandafter \@firstoftwo
 \else \expandafter \@secondoftwo
 \fi
}%
\providecommand \natexlab [1]{#1}%
\providecommand \enquote  [1]{``#1''}%
\providecommand \bibnamefont  [1]{#1}%
\providecommand \bibfnamefont [1]{#1}%
\providecommand \citenamefont [1]{#1}%
\providecommand \href@noop [0]{\@secondoftwo}%
\providecommand \href [0]{\begingroup \@sanitize@url \@href}%
\providecommand \@href[1]{\@@startlink{#1}\@@href}%
\providecommand \@@href[1]{\endgroup#1\@@endlink}%
\providecommand \@sanitize@url [0]{\catcode `\\12\catcode `\$12\catcode
  `\&12\catcode `\#12\catcode `\^12\catcode `\_12\catcode `\%12\relax}%
\providecommand \@@startlink[1]{}%
\providecommand \@@endlink[0]{}%
\providecommand \url  [0]{\begingroup\@sanitize@url \@url }%
\providecommand \@url [1]{\endgroup\@href {#1}{\urlprefix }}%
\providecommand \urlprefix  [0]{URL }%
\providecommand \Eprint [0]{\href }%
\providecommand \doibase [0]{http://dx.doi.org/}%
\providecommand \selectlanguage [0]{\@gobble}%
\providecommand \bibinfo  [0]{\@secondoftwo}%
\providecommand \bibfield  [0]{\@secondoftwo}%
\providecommand \translation [1]{[#1]}%
\providecommand \BibitemOpen [0]{}%
\providecommand \bibitemStop [0]{}%
\providecommand \bibitemNoStop [0]{.\EOS\space}%
\providecommand \EOS [0]{\spacefactor3000\relax}%
\providecommand \BibitemShut  [1]{\csname bibitem#1\endcsname}%
\let\auto@bib@innerbib\@empty
\bibitem [{\citenamefont {Pal}\ and\ \citenamefont
  {Wolfenstein}(1982)}]{Pal:1981rm}%
  \BibitemOpen
  \bibfield  {author} {\bibinfo {author} {\bibfnamefont {P.~B.}\ \bibnamefont
  {Pal}}\ and\ \bibinfo {author} {\bibfnamefont {L.}~\bibnamefont
  {Wolfenstein}},\ }\href {\doibase 10.1103/PhysRevD.25.766} {\bibfield
  {journal} {\bibinfo  {journal} {Phys.~Rev.~D}\ }\textbf {\bibinfo {volume}
  {25}},\ \bibinfo {pages} {766} (\bibinfo {year} {1982})}\BibitemShut
  {NoStop}%
\bibitem [{\citenamefont {Hosotani}(1981)}]{Hosotani:1981mq}%
  \BibitemOpen
  \bibfield  {author} {\bibinfo {author} {\bibfnamefont {Y.}~\bibnamefont
  {Hosotani}},\ }\href {\doibase 10.1016/0550-3213(81)90306-0} {\bibfield
  {journal} {\bibinfo  {journal} {Nucl.~Phys.~B}\ }\textbf {\bibinfo {volume}
  {191}},\ \bibinfo {pages} {411} (\bibinfo {year} {1981})}\BibitemShut
  {NoStop}%
\bibitem [{\citenamefont {Nieves}(1983)}]{Nieves:1982bq}%
  \BibitemOpen
  \bibfield  {author} {\bibinfo {author} {\bibfnamefont {J.~F.}\ \bibnamefont
  {Nieves}},\ }\href {\doibase 10.1103/PhysRevD.28.1664} {\bibfield  {journal}
  {\bibinfo  {journal} {Phys.~Rev.~D}\ }\textbf {\bibinfo {volume} {28}},\
  \bibinfo {pages} {1664} (\bibinfo {year} {1983})}\BibitemShut {NoStop}%
\bibitem [{\citenamefont {Bahcall}\ \emph {et~al.}(1972)\citenamefont
  {Bahcall}, \citenamefont {Cabibbo},\ and\ \citenamefont
  {Yahil}}]{Bahcall:1972my}%
  \BibitemOpen
  \bibfield  {author} {\bibinfo {author} {\bibfnamefont {J.~N.}\ \bibnamefont
  {Bahcall}}, \bibinfo {author} {\bibfnamefont {N.}~\bibnamefont {Cabibbo}}, \
  and\ \bibinfo {author} {\bibfnamefont {A.}~\bibnamefont {Yahil}},\ }\href
  {\doibase 10.1103/PhysRevLett.28.316} {\bibfield  {journal} {\bibinfo
  {journal} {Phys.\ Rev.\ Lett.}\ }\textbf {\bibinfo {volume} {28}},\ \bibinfo
  {pages} {316} (\bibinfo {year} {1972})}\BibitemShut {NoStop}%
\bibitem [{\citenamefont {Chikashige}\ \emph {et~al.}(1980)\citenamefont
  {Chikashige}, \citenamefont {Mohapatra},\ and\ \citenamefont
  {Peccei}}]{Chikashige:1980qk}%
  \BibitemOpen
  \bibfield  {author} {\bibinfo {author} {\bibfnamefont {Y.}~\bibnamefont
  {Chikashige}}, \bibinfo {author} {\bibfnamefont {R.~N.}\ \bibnamefont
  {Mohapatra}}, \ and\ \bibinfo {author} {\bibfnamefont {R.~D.}\ \bibnamefont
  {Peccei}},\ }\href {\doibase 10.1103/PhysRevLett.45.1926} {\bibfield
  {journal} {\bibinfo  {journal} {Phys.~Rev.~Lett.}\ }\textbf {\bibinfo
  {volume} {45}},\ \bibinfo {pages} {1926} (\bibinfo {year}
  {1980})}\BibitemShut {NoStop}%
\bibitem [{\citenamefont {Gelmini}\ \emph {et~al.}(1982)\citenamefont
  {Gelmini}, \citenamefont {Nussinov},\ and\ \citenamefont
  {Roncadelli}}]{Gelmini:1982rr}%
  \BibitemOpen
  \bibfield  {author} {\bibinfo {author} {\bibfnamefont {G.~B.}\ \bibnamefont
  {Gelmini}}, \bibinfo {author} {\bibfnamefont {S.}~\bibnamefont {Nussinov}}, \
  and\ \bibinfo {author} {\bibfnamefont {M.}~\bibnamefont {Roncadelli}},\
  }\href {\doibase 10.1016/0550-3213(82)90107-9} {\bibfield  {journal}
  {\bibinfo  {journal} {Nucl.~Phys.~B}\ }\textbf {\bibinfo {volume} {209}},\
  \bibinfo {pages} {157} (\bibinfo {year} {1982})}\BibitemShut {NoStop}%
\bibitem [{\citenamefont {Tomas}\ \emph {et~al.}(2001)\citenamefont {Tomas},
  \citenamefont {Pas},\ and\ \citenamefont {Valle}}]{Tomas:2001dh}%
  \BibitemOpen
  \bibfield  {author} {\bibinfo {author} {\bibfnamefont {R.}~\bibnamefont
  {Tomas}}, \bibinfo {author} {\bibfnamefont {H.}~\bibnamefont {Pas}}, \ and\
  \bibinfo {author} {\bibfnamefont {J.~W.~F.}\ \bibnamefont {Valle}},\ }\href
  {\doibase 10.1103/PhysRevD.64.095005} {\bibfield  {journal} {\bibinfo
  {journal} {Phys.~Rev.~D}\ }\textbf {\bibinfo {volume} {64}},\ \bibinfo
  {pages} {095005} (\bibinfo {year} {2001})},\ \Eprint
  {http://arxiv.org/abs/hep-ph/0103017} {arXiv:hep-ph/0103017} \BibitemShut
  {NoStop}%
\bibitem [{\citenamefont {Hannestad}\ and\ \citenamefont
  {Raffelt}(2005)}]{Hannestad:2005ex}%
  \BibitemOpen
  \bibfield  {author} {\bibinfo {author} {\bibfnamefont {S.}~\bibnamefont
  {Hannestad}}\ and\ \bibinfo {author} {\bibfnamefont {G.}~\bibnamefont
  {Raffelt}},\ }\href {\doibase 10.1103/PhysRevD.72.103514} {\bibfield
  {journal} {\bibinfo  {journal} {Phys.~Rev.~D}\ }\textbf {\bibinfo {volume}
  {72}},\ \bibinfo {pages} {103514} (\bibinfo {year} {2005})},\ \Eprint
  {http://arxiv.org/abs/hep-ph/0509278} {arXiv:hep-ph/0509278 [hep-ph]}
  \BibitemShut {NoStop}%
\bibitem [{\citenamefont {Zhou}(2008)}]{Zhou:2007zq}%
  \BibitemOpen
  \bibfield  {author} {\bibinfo {author} {\bibfnamefont {S.}~\bibnamefont
  {Zhou}},\ }\href {\doibase 10.1016/j.physletb.2007.10.081} {\bibfield
  {journal} {\bibinfo  {journal} {Phys.~Lett.~B}\ }\textbf {\bibinfo {volume}
  {659}},\ \bibinfo {pages} {336} (\bibinfo {year} {2008})},\ \Eprint
  {http://arxiv.org/abs/0706.0302} {arXiv:0706.0302 [hep-ph]} \BibitemShut
  {NoStop}%
\bibitem [{\citenamefont {Chen}\ \emph {et~al.}(2007)\citenamefont {Chen},
  \citenamefont {He},\ and\ \citenamefont {Tsai}}]{Chen:2007zy}%
  \BibitemOpen
  \bibfield  {author} {\bibinfo {author} {\bibfnamefont {S.-L.}\ \bibnamefont
  {Chen}}, \bibinfo {author} {\bibfnamefont {X.-G.}\ \bibnamefont {He}}, \ and\
  \bibinfo {author} {\bibfnamefont {H.-C.}\ \bibnamefont {Tsai}},\ }\href
  {\doibase 10.1088/1126-6708/2007/11/010} {\bibfield  {journal} {\bibinfo
  {journal} {JHEP}\ }\textbf {\bibinfo {volume} {11}},\ \bibinfo {pages} {010}
  (\bibinfo {year} {2007})},\ \Eprint {http://arxiv.org/abs/0707.0187}
  {arXiv:0707.0187 [hep-ph]} \BibitemShut {NoStop}%
\bibitem [{\citenamefont {Li}\ \emph {et~al.}(2008)\citenamefont {Li},
  \citenamefont {Liu},\ and\ \citenamefont {Wei}}]{Li:2007kj}%
  \BibitemOpen
  \bibfield  {author} {\bibinfo {author} {\bibfnamefont {X.-Q.}\ \bibnamefont
  {Li}}, \bibinfo {author} {\bibfnamefont {Y.}~\bibnamefont {Liu}}, \ and\
  \bibinfo {author} {\bibfnamefont {Z.-T.}\ \bibnamefont {Wei}},\ }\href
  {\doibase 10.1140/epjc/s10052-008-0630-6} {\bibfield  {journal} {\bibinfo
  {journal} {Eur.~Phys.~J.~C}\ }\textbf {\bibinfo {volume} {56}},\ \bibinfo
  {pages} {97} (\bibinfo {year} {2008})},\ \Eprint
  {http://arxiv.org/abs/0707.2285} {arXiv:0707.2285 [hep-ph]} \BibitemShut
  {NoStop}%
\bibitem [{\citenamefont {Escudero}\ and\ \citenamefont
  {Fairbairn}(2019)}]{Escudero:2019gfk}%
  \BibitemOpen
  \bibfield  {author} {\bibinfo {author} {\bibfnamefont {M.}~\bibnamefont
  {Escudero}}\ and\ \bibinfo {author} {\bibfnamefont {M.}~\bibnamefont
  {Fairbairn}},\ }\href {\doibase 10.1103/PhysRevD.100.103531} {\bibfield
  {journal} {\bibinfo  {journal} {Phys.\ Rev.\ D}\ }\textbf {\bibinfo {volume}
  {100}},\ \bibinfo {pages} {103531} (\bibinfo {year} {2019})},\ \Eprint
  {http://arxiv.org/abs/1907.05425} {arXiv:1907.05425 [hep-ph]} \BibitemShut
  {NoStop}%
\bibitem [{\citenamefont {Aartsen}\ \emph
  {et~al.}(2013{\natexlab{a}})\citenamefont {Aartsen} \emph
  {et~al.}}]{Aartsen:2013bka}%
  \BibitemOpen
  \bibfield  {author} {\bibinfo {author} {\bibfnamefont {M.~G.}\ \bibnamefont
  {Aartsen}} \emph {et~al.} (\bibinfo {collaboration} {IceCube}),\ }\href
  {\doibase 10.1103/PhysRevLett.111.021103} {\bibfield  {journal} {\bibinfo
  {journal} {Phys.~Rev.~Lett.}\ }\textbf {\bibinfo {volume} {111}},\ \bibinfo
  {pages} {021103} (\bibinfo {year} {2013}{\natexlab{a}})},\ \Eprint
  {http://arxiv.org/abs/1304.5356} {arXiv:1304.5356 [astro-ph.HE]} \BibitemShut
  {NoStop}%
\bibitem [{\citenamefont {Aartsen}\ \emph
  {et~al.}(2013{\natexlab{b}})\citenamefont {Aartsen} \emph
  {et~al.}}]{Aartsen:2013jdh}%
  \BibitemOpen
  \bibfield  {author} {\bibinfo {author} {\bibfnamefont {M.~G.}\ \bibnamefont
  {Aartsen}} \emph {et~al.} (\bibinfo {collaboration} {IceCube}),\ }\href
  {\doibase 10.1126/science.1242856} {\bibfield  {journal} {\bibinfo  {journal}
  {Science}\ }\textbf {\bibinfo {volume} {342}},\ \bibinfo {pages} {1242856}
  (\bibinfo {year} {2013}{\natexlab{b}})},\ \Eprint
  {http://arxiv.org/abs/1311.5238} {arXiv:1311.5238 [astro-ph.HE]} \BibitemShut
  {NoStop}%
\bibitem [{\citenamefont {Aartsen}\ \emph
  {et~al.}(2014{\natexlab{a}})\citenamefont {Aartsen} \emph
  {et~al.}}]{Aartsen:2014gkd}%
  \BibitemOpen
  \bibfield  {author} {\bibinfo {author} {\bibfnamefont {M.~G.}\ \bibnamefont
  {Aartsen}} \emph {et~al.} (\bibinfo {collaboration} {IceCube}),\ }\href
  {\doibase 10.1103/PhysRevLett.113.101101} {\bibfield  {journal} {\bibinfo
  {journal} {Phys.\ Rev.\ Lett.}\ }\textbf {\bibinfo {volume} {113}},\ \bibinfo
  {pages} {101101} (\bibinfo {year} {2014}{\natexlab{a}})},\ \Eprint
  {http://arxiv.org/abs/1405.5303} {arXiv:1405.5303 [astro-ph.HE]} \BibitemShut
  {NoStop}%
\bibitem [{\citenamefont {Aartsen}\ \emph {et~al.}(2015)\citenamefont {Aartsen}
  \emph {et~al.}}]{Aartsen:2015rwa}%
  \BibitemOpen
  \bibfield  {author} {\bibinfo {author} {\bibfnamefont {M.~G.}\ \bibnamefont
  {Aartsen}} \emph {et~al.} (\bibinfo {collaboration} {IceCube}),\ }\href
  {\doibase 10.1103/PhysRevLett.115.081102} {\bibfield  {journal} {\bibinfo
  {journal} {Phys.~Rev.~Lett.}\ }\textbf {\bibinfo {volume} {115}},\ \bibinfo
  {pages} {081102} (\bibinfo {year} {2015})},\ \Eprint
  {http://arxiv.org/abs/1507.04005} {arXiv:1507.04005 [astro-ph.HE]}
  \BibitemShut {NoStop}%
\bibitem [{\citenamefont {Aartsen}\ \emph {et~al.}(2016)\citenamefont {Aartsen}
  \emph {et~al.}}]{Aartsen:2016xlq}%
  \BibitemOpen
  \bibfield  {author} {\bibinfo {author} {\bibfnamefont {M.~G.}\ \bibnamefont
  {Aartsen}} \emph {et~al.} (\bibinfo {collaboration} {IceCube}),\ }\href
  {\doibase 10.3847/0004-637X/833/1/3} {\bibfield  {journal} {\bibinfo
  {journal} {Astrophys.\ J.}\ }\textbf {\bibinfo {volume} {833}},\ \bibinfo
  {pages} {3} (\bibinfo {year} {2016})},\ \Eprint
  {http://arxiv.org/abs/1607.08006} {arXiv:1607.08006 [astro-ph.HE]}
  \BibitemShut {NoStop}%
\bibitem [{\citenamefont {Pakvasa}(1981)}]{Pakvasa:1981ci}%
  \BibitemOpen
  \bibfield  {author} {\bibinfo {author} {\bibfnamefont {S.}~\bibnamefont
  {Pakvasa}},\ }\href@noop {} {\bibfield  {journal} {\bibinfo  {journal} {Nuovo
  Cim.\ Lett.}\ }\textbf {\bibinfo {volume} {31}},\ \bibinfo {pages} {497}
  (\bibinfo {year} {1981})}\BibitemShut {NoStop}%
\bibitem [{\citenamefont {Beacom}\ \emph
  {et~al.}(2003{\natexlab{a}})\citenamefont {Beacom}, \citenamefont {Bell},
  \citenamefont {Hooper}, \citenamefont {Pakvasa},\ and\ \citenamefont
  {Weiler}}]{Beacom:2002vi}%
  \BibitemOpen
  \bibfield  {author} {\bibinfo {author} {\bibfnamefont {J.~F.}\ \bibnamefont
  {Beacom}}, \bibinfo {author} {\bibfnamefont {N.~F.}\ \bibnamefont {Bell}},
  \bibinfo {author} {\bibfnamefont {D.}~\bibnamefont {Hooper}}, \bibinfo
  {author} {\bibfnamefont {S.}~\bibnamefont {Pakvasa}}, \ and\ \bibinfo
  {author} {\bibfnamefont {T.~J.}\ \bibnamefont {Weiler}},\ }\href {\doibase
  10.1103/PhysRevLett.90.181301} {\bibfield  {journal} {\bibinfo  {journal}
  {Phys.~Rev.~Lett.}\ }\textbf {\bibinfo {volume} {90}},\ \bibinfo {pages}
  {181301} (\bibinfo {year} {2003}{\natexlab{a}})},\ \Eprint
  {http://arxiv.org/abs/hep-ph/0211305} {arXiv:hep-ph/0211305} \BibitemShut
  {NoStop}%
\bibitem [{\citenamefont {Barenboim}\ and\ \citenamefont
  {Quigg}(2003)}]{Barenboim:2003jm}%
  \BibitemOpen
  \bibfield  {author} {\bibinfo {author} {\bibfnamefont {G.}~\bibnamefont
  {Barenboim}}\ and\ \bibinfo {author} {\bibfnamefont {C.}~\bibnamefont
  {Quigg}},\ }\href {\doibase 10.1103/PhysRevD.67.073024} {\bibfield  {journal}
  {\bibinfo  {journal} {Phys.~Rev.~D}\ }\textbf {\bibinfo {volume} {67}},\
  \bibinfo {pages} {073024} (\bibinfo {year} {2003})},\ \Eprint
  {http://arxiv.org/abs/hep-ph/0301220} {arXiv:hep-ph/0301220} \BibitemShut
  {NoStop}%
\bibitem [{\citenamefont {Beacom}\ \emph
  {et~al.}(2003{\natexlab{b}})\citenamefont {Beacom}, \citenamefont {Bell},
  \citenamefont {Hooper}, \citenamefont {Pakvasa},\ and\ \citenamefont
  {Weiler}}]{Beacom:2003nh}%
  \BibitemOpen
  \bibfield  {author} {\bibinfo {author} {\bibfnamefont {J.~F.}\ \bibnamefont
  {Beacom}}, \bibinfo {author} {\bibfnamefont {N.~F.}\ \bibnamefont {Bell}},
  \bibinfo {author} {\bibfnamefont {D.}~\bibnamefont {Hooper}}, \bibinfo
  {author} {\bibfnamefont {S.}~\bibnamefont {Pakvasa}}, \ and\ \bibinfo
  {author} {\bibfnamefont {T.~J.}\ \bibnamefont {Weiler}},\ }\href {\doibase
  10.1103/PhysRevD.72.019901} {\bibfield  {journal} {\bibinfo  {journal}
  {Phys.\ Rev.\ D}\ }\textbf {\bibinfo {volume} {68}},\ \bibinfo {pages}
  {093005} (\bibinfo {year} {2003}{\natexlab{b}})},\ \bibinfo {note} {[Erratum:
  Phys.\ Rev.\ D 72, 019901 (2005)]},\ \Eprint
  {http://arxiv.org/abs/hep-ph/0307025} {arXiv:hep-ph/0307025} \BibitemShut
  {NoStop}%
\bibitem [{\citenamefont {Beacom}\ \emph
  {et~al.}(2004{\natexlab{a}})\citenamefont {Beacom}, \citenamefont {Bell},
  \citenamefont {Hooper}, \citenamefont {Pakvasa},\ and\ \citenamefont
  {Weiler}}]{Beacom:2003zg}%
  \BibitemOpen
  \bibfield  {author} {\bibinfo {author} {\bibfnamefont {J.~F.}\ \bibnamefont
  {Beacom}}, \bibinfo {author} {\bibfnamefont {N.~F.}\ \bibnamefont {Bell}},
  \bibinfo {author} {\bibfnamefont {D.}~\bibnamefont {Hooper}}, \bibinfo
  {author} {\bibfnamefont {S.}~\bibnamefont {Pakvasa}}, \ and\ \bibinfo
  {author} {\bibfnamefont {T.~J.}\ \bibnamefont {Weiler}},\ }\href@noop {}
  {\bibfield  {journal} {\bibinfo  {journal} {Phys.~Rev.~D}\ }\textbf {\bibinfo
  {volume} {69}},\ \bibinfo {pages} {017303} (\bibinfo {year}
  {2004}{\natexlab{a}})},\ \Eprint {http://arxiv.org/abs/hep-ph/0309267}
  {hep-ph/0309267} \BibitemShut {NoStop}%
\bibitem [{\citenamefont {Meloni}\ and\ \citenamefont
  {Ohlsson}(2007)}]{Meloni:2006gv}%
  \BibitemOpen
  \bibfield  {author} {\bibinfo {author} {\bibfnamefont {D.}~\bibnamefont
  {Meloni}}\ and\ \bibinfo {author} {\bibfnamefont {T.}~\bibnamefont
  {Ohlsson}},\ }\href {\doibase 10.1103/PhysRevD.75.125017} {\bibfield
  {journal} {\bibinfo  {journal} {Phys.\ Rev.\ D}\ }\textbf {\bibinfo {volume}
  {75}},\ \bibinfo {pages} {125017} (\bibinfo {year} {2007})},\ \Eprint
  {http://arxiv.org/abs/hep-ph/0612279} {arXiv:hep-ph/0612279} \BibitemShut
  {NoStop}%
\bibitem [{\citenamefont {Maltoni}\ and\ \citenamefont
  {Winter}(2008)}]{Maltoni:2008jr}%
  \BibitemOpen
  \bibfield  {author} {\bibinfo {author} {\bibfnamefont {M.}~\bibnamefont
  {Maltoni}}\ and\ \bibinfo {author} {\bibfnamefont {W.}~\bibnamefont
  {Winter}},\ }\href {\doibase 10.1088/1126-6708/2008/07/064} {\bibfield
  {journal} {\bibinfo  {journal} {JHEP}\ }\textbf {\bibinfo {volume} {07}},\
  \bibinfo {pages} {064} (\bibinfo {year} {2008})},\ \Eprint
  {http://arxiv.org/abs/0803.2050} {arXiv:0803.2050 [hep-ph]} \BibitemShut
  {NoStop}%
\bibitem [{\citenamefont {Bustamante}\ \emph {et~al.}(2010)\citenamefont
  {Bustamante}, \citenamefont {Gago},\ and\ \citenamefont {Pe\~na
  Garay}}]{Bustamante:2010nq}%
  \BibitemOpen
  \bibfield  {author} {\bibinfo {author} {\bibfnamefont {M.}~\bibnamefont
  {Bustamante}}, \bibinfo {author} {\bibfnamefont {A.~M.}\ \bibnamefont
  {Gago}}, \ and\ \bibinfo {author} {\bibfnamefont {C.}~\bibnamefont {Pe\~na
  Garay}},\ }\href {\doibase 10.1007/JHEP04(2010)066} {\bibfield  {journal}
  {\bibinfo  {journal} {JHEP}\ }\textbf {\bibinfo {volume} {1004}},\ \bibinfo
  {pages} {066} (\bibinfo {year} {2010})},\ \Eprint
  {http://arxiv.org/abs/1001.4878} {arXiv:1001.4878 [hep-ph]} \BibitemShut
  {NoStop}%
\bibitem [{\citenamefont {Mehta}\ and\ \citenamefont
  {Winter}(2011)}]{Mehta:2011qb}%
  \BibitemOpen
  \bibfield  {author} {\bibinfo {author} {\bibfnamefont {P.}~\bibnamefont
  {Mehta}}\ and\ \bibinfo {author} {\bibfnamefont {W.}~\bibnamefont {Winter}},\
  }\href@noop {} {\bibfield  {journal} {\bibinfo  {journal} {JCAP}\ }\textbf
  {\bibinfo {volume} {1103}},\ \bibinfo {pages} {041} (\bibinfo {year}
  {2011})},\ \Eprint {http://arxiv.org/abs/1101.2673} {arXiv:1101.2673
  [hep-ph]} \BibitemShut {NoStop}%
\bibitem [{\citenamefont {Baerwald}\ \emph {et~al.}(2012)\citenamefont
  {Baerwald}, \citenamefont {Bustamante},\ and\ \citenamefont
  {Winter}}]{Baerwald:2012kc}%
  \BibitemOpen
  \bibfield  {author} {\bibinfo {author} {\bibfnamefont {P.}~\bibnamefont
  {Baerwald}}, \bibinfo {author} {\bibfnamefont {M.}~\bibnamefont
  {Bustamante}}, \ and\ \bibinfo {author} {\bibfnamefont {W.}~\bibnamefont
  {Winter}},\ }\href {\doibase 10.1088/1475-7516/2012/10/020} {\bibfield
  {journal} {\bibinfo  {journal} {JCAP}\ }\textbf {\bibinfo {volume} {1210}},\
  \bibinfo {pages} {020} (\bibinfo {year} {2012})},\ \Eprint
  {http://arxiv.org/abs/1208.4600} {arXiv:1208.4600 [astro-ph.CO]} \BibitemShut
  {NoStop}%
\bibitem [{\citenamefont {Pakvasa}\ \emph {et~al.}(2013)\citenamefont
  {Pakvasa}, \citenamefont {Joshipura},\ and\ \citenamefont
  {Mohanty}}]{Pakvasa:2012db}%
  \BibitemOpen
  \bibfield  {author} {\bibinfo {author} {\bibfnamefont {S.}~\bibnamefont
  {Pakvasa}}, \bibinfo {author} {\bibfnamefont {A.}~\bibnamefont {Joshipura}},
  \ and\ \bibinfo {author} {\bibfnamefont {S.}~\bibnamefont {Mohanty}},\ }\href
  {\doibase 10.1103/PhysRevLett.110.171802} {\bibfield  {journal} {\bibinfo
  {journal} {Phys.\ Rev.\ Lett.}\ }\textbf {\bibinfo {volume} {110}},\ \bibinfo
  {pages} {171802} (\bibinfo {year} {2013})},\ \Eprint
  {http://arxiv.org/abs/1209.5630} {arXiv:1209.5630 [hep-ph]} \BibitemShut
  {NoStop}%
\bibitem [{\citenamefont {Pagliaroli}\ \emph {et~al.}(2015)\citenamefont
  {Pagliaroli}, \citenamefont {Palladino}, \citenamefont {Villante},\ and\
  \citenamefont {Vissani}}]{Pagliaroli:2015rca}%
  \BibitemOpen
  \bibfield  {author} {\bibinfo {author} {\bibfnamefont {G.}~\bibnamefont
  {Pagliaroli}}, \bibinfo {author} {\bibfnamefont {A.}~\bibnamefont
  {Palladino}}, \bibinfo {author} {\bibfnamefont {F.~L.}\ \bibnamefont
  {Villante}}, \ and\ \bibinfo {author} {\bibfnamefont {F.}~\bibnamefont
  {Vissani}},\ }\href {\doibase 10.1103/PhysRevD.92.113008} {\bibfield
  {journal} {\bibinfo  {journal} {Phys.\ Rev.\ D}\ }\textbf {\bibinfo {volume}
  {92}},\ \bibinfo {pages} {113008} (\bibinfo {year} {2015})},\ \Eprint
  {http://arxiv.org/abs/1506.02624} {arXiv:1506.02624 [hep-ph]} \BibitemShut
  {NoStop}%
\bibitem [{\citenamefont {Bustamante}\ \emph {et~al.}(2015)\citenamefont
  {Bustamante}, \citenamefont {Beacom},\ and\ \citenamefont
  {Winter}}]{Bustamante:2015waa}%
  \BibitemOpen
  \bibfield  {author} {\bibinfo {author} {\bibfnamefont {M.}~\bibnamefont
  {Bustamante}}, \bibinfo {author} {\bibfnamefont {J.~F.}\ \bibnamefont
  {Beacom}}, \ and\ \bibinfo {author} {\bibfnamefont {W.}~\bibnamefont
  {Winter}},\ }\href {\doibase 10.1103/PhysRevLett.115.161302} {\bibfield
  {journal} {\bibinfo  {journal} {Phys.\ Rev.\ Lett.}\ }\textbf {\bibinfo
  {volume} {115}},\ \bibinfo {pages} {161302} (\bibinfo {year} {2015})},\
  \Eprint {http://arxiv.org/abs/1506.02645} {arXiv:1506.02645 [astro-ph.HE]}
  \BibitemShut {NoStop}%
\bibitem [{\citenamefont {Huang}\ and\ \citenamefont
  {Ma}(2015)}]{Huang:2015flc}%
  \BibitemOpen
  \bibfield  {author} {\bibinfo {author} {\bibfnamefont {Y.}~\bibnamefont
  {Huang}}\ and\ \bibinfo {author} {\bibfnamefont {B.-Q.}\ \bibnamefont {Ma}},\
  }\href@noop {} {\bibfield  {journal} {\bibinfo  {journal} {The~Universe}\
  }\textbf {\bibinfo {volume} {3}},\ \bibinfo {pages} {15} (\bibinfo {year}
  {2015})},\ \Eprint {http://arxiv.org/abs/1508.01698} {arXiv:1508.01698
  [hep-ph]} \BibitemShut {NoStop}%
\bibitem [{\citenamefont {Shoemaker}\ and\ \citenamefont
  {Murase}(2016)}]{Shoemaker:2015qul}%
  \BibitemOpen
  \bibfield  {author} {\bibinfo {author} {\bibfnamefont {I.~M.}\ \bibnamefont
  {Shoemaker}}\ and\ \bibinfo {author} {\bibfnamefont {K.}~\bibnamefont
  {Murase}},\ }\href {\doibase 10.1103/PhysRevD.93.085004} {\bibfield
  {journal} {\bibinfo  {journal} {Phys.\ Rev.\ D}\ }\textbf {\bibinfo {volume}
  {93}},\ \bibinfo {pages} {085004} (\bibinfo {year} {2016})},\ \Eprint
  {http://arxiv.org/abs/1512.07228} {arXiv:1512.07228 [astro-ph.HE]}
  \BibitemShut {NoStop}%
\bibitem [{\citenamefont {Bustamante}\ \emph {et~al.}(2017)\citenamefont
  {Bustamante}, \citenamefont {Beacom},\ and\ \citenamefont
  {Murase}}]{Bustamante:2016ciw}%
  \BibitemOpen
  \bibfield  {author} {\bibinfo {author} {\bibfnamefont {M.}~\bibnamefont
  {Bustamante}}, \bibinfo {author} {\bibfnamefont {J.~F.}\ \bibnamefont
  {Beacom}}, \ and\ \bibinfo {author} {\bibfnamefont {K.}~\bibnamefont
  {Murase}},\ }\href {\doibase 10.1103/PhysRevD.95.063013} {\bibfield
  {journal} {\bibinfo  {journal} {Phys.\ Rev.\ D}\ }\textbf {\bibinfo {volume}
  {95}},\ \bibinfo {pages} {063013} (\bibinfo {year} {2017})},\ \Eprint
  {http://arxiv.org/abs/1610.02096} {arXiv:1610.02096 [astro-ph.HE]}
  \BibitemShut {NoStop}%
\bibitem [{\citenamefont {Rasmussen}\ \emph {et~al.}(2017)\citenamefont
  {Rasmussen}, \citenamefont {Lechner}, \citenamefont {Ackermann},
  \citenamefont {Kowalski},\ and\ \citenamefont {Winter}}]{Rasmussen:2017ert}%
  \BibitemOpen
  \bibfield  {author} {\bibinfo {author} {\bibfnamefont {R.~W.}\ \bibnamefont
  {Rasmussen}}, \bibinfo {author} {\bibfnamefont {L.}~\bibnamefont {Lechner}},
  \bibinfo {author} {\bibfnamefont {M.}~\bibnamefont {Ackermann}}, \bibinfo
  {author} {\bibfnamefont {M.}~\bibnamefont {Kowalski}}, \ and\ \bibinfo
  {author} {\bibfnamefont {W.}~\bibnamefont {Winter}},\ }\href {\doibase
  10.1103/PhysRevD.96.083018} {\bibfield  {journal} {\bibinfo  {journal}
  {Phys.\ Rev.\ D}\ }\textbf {\bibinfo {volume} {96}},\ \bibinfo {pages}
  {083018} (\bibinfo {year} {2017})},\ \Eprint
  {http://arxiv.org/abs/1707.07684} {arXiv:1707.07684 [hep-ph]} \BibitemShut
  {NoStop}%
\bibitem [{\citenamefont {Ahlers}\ \emph {et~al.}(2018)\citenamefont {Ahlers},
  \citenamefont {Helbing},\ and\ \citenamefont {P\'erez de~los
  Heros}}]{Ahlers:2018mkf}%
  \BibitemOpen
  \bibfield  {author} {\bibinfo {author} {\bibfnamefont {M.}~\bibnamefont
  {Ahlers}}, \bibinfo {author} {\bibfnamefont {K.}~\bibnamefont {Helbing}}, \
  and\ \bibinfo {author} {\bibfnamefont {C.}~\bibnamefont {P\'erez de~los
  Heros}},\ }\href {\doibase 10.1140/epjc/s10052-018-6369-9} {\bibfield
  {journal} {\bibinfo  {journal} {Eur.\ Phys.\ J.\ C}\ }\textbf {\bibinfo
  {volume} {78}},\ \bibinfo {pages} {924} (\bibinfo {year} {2018})},\ \Eprint
  {http://arxiv.org/abs/1806.05696} {arXiv:1806.05696 [astro-ph.HE]}
  \BibitemShut {NoStop}%
\bibitem [{\citenamefont {Denton}\ and\ \citenamefont
  {Tamborra}(2018)}]{Denton:2018aml}%
  \BibitemOpen
  \bibfield  {author} {\bibinfo {author} {\bibfnamefont {P.~B.}\ \bibnamefont
  {Denton}}\ and\ \bibinfo {author} {\bibfnamefont {I.}~\bibnamefont
  {Tamborra}},\ }\href {\doibase 10.1103/PhysRevLett.121.121802} {\bibfield
  {journal} {\bibinfo  {journal} {Phys.\ Rev.\ Lett.}\ }\textbf {\bibinfo
  {volume} {121}},\ \bibinfo {pages} {121802} (\bibinfo {year} {2018})},\
  \Eprint {http://arxiv.org/abs/1805.05950} {arXiv:1805.05950 [hep-ph]}
  \BibitemShut {NoStop}%
\bibitem [{\citenamefont {Berryman}\ \emph {et~al.}(2015)\citenamefont
  {Berryman}, \citenamefont {de~Gouvea},\ and\ \citenamefont
  {Hernandez}}]{Berryman:2014qha}%
  \BibitemOpen
  \bibfield  {author} {\bibinfo {author} {\bibfnamefont {J.~M.}\ \bibnamefont
  {Berryman}}, \bibinfo {author} {\bibfnamefont {A.}~\bibnamefont {de~Gouvea}},
  \ and\ \bibinfo {author} {\bibfnamefont {D.}~\bibnamefont {Hernandez}},\
  }\href {\doibase 10.1103/PhysRevD.92.073003} {\bibfield  {journal} {\bibinfo
  {journal} {Phys.~Rev.~D}\ }\textbf {\bibinfo {volume} {92}},\ \bibinfo
  {pages} {073003} (\bibinfo {year} {2015})},\ \Eprint
  {http://arxiv.org/abs/1411.0308} {arXiv:1411.0308 [hep-ph]} \BibitemShut
  {NoStop}%
\bibitem [{\citenamefont {Joshipura}\ \emph {et~al.}(2002)\citenamefont
  {Joshipura}, \citenamefont {Masso},\ and\ \citenamefont
  {Mohanty}}]{Joshipura:2002fb}%
  \BibitemOpen
  \bibfield  {author} {\bibinfo {author} {\bibfnamefont {A.~S.}\ \bibnamefont
  {Joshipura}}, \bibinfo {author} {\bibfnamefont {E.}~\bibnamefont {Masso}}, \
  and\ \bibinfo {author} {\bibfnamefont {S.}~\bibnamefont {Mohanty}},\ }\href
  {\doibase 10.1103/PhysRevD.66.113008} {\bibfield  {journal} {\bibinfo
  {journal} {Phys.\ Rev.\ D}\ }\textbf {\bibinfo {volume} {66}},\ \bibinfo
  {pages} {113008} (\bibinfo {year} {2002})},\ \Eprint
  {http://arxiv.org/abs/hep-ph/0203181} {arXiv:hep-ph/0203181} \BibitemShut
  {NoStop}%
\bibitem [{\citenamefont {Beacom}\ and\ \citenamefont
  {Bell}(2002)}]{Beacom:2002cb}%
  \BibitemOpen
  \bibfield  {author} {\bibinfo {author} {\bibfnamefont {J.~F.}\ \bibnamefont
  {Beacom}}\ and\ \bibinfo {author} {\bibfnamefont {N.~F.}\ \bibnamefont
  {Bell}},\ }\href {\doibase 10.1103/PhysRevD.65.113009} {\bibfield  {journal}
  {\bibinfo  {journal} {Phys.\ Rev.\ D}\ }\textbf {\bibinfo {volume} {65}},\
  \bibinfo {pages} {113009} (\bibinfo {year} {2002})},\ \Eprint
  {http://arxiv.org/abs/hep-ph/0204111} {arXiv:hep-ph/0204111 [hep-ph]}
  \BibitemShut {NoStop}%
\bibitem [{\citenamefont {Bandyopadhyay}\ \emph {et~al.}(2002)\citenamefont
  {Bandyopadhyay}, \citenamefont {Choubey}, \citenamefont {Goswami},\ and\
  \citenamefont {Roy}}]{Bandyopadhyay:2002xj}%
  \BibitemOpen
  \bibfield  {author} {\bibinfo {author} {\bibfnamefont {A.}~\bibnamefont
  {Bandyopadhyay}}, \bibinfo {author} {\bibfnamefont {S.}~\bibnamefont
  {Choubey}}, \bibinfo {author} {\bibfnamefont {S.}~\bibnamefont {Goswami}}, \
  and\ \bibinfo {author} {\bibfnamefont {D.~P.}\ \bibnamefont {Roy}},\ }\href
  {\doibase 10.1016/S0370-2693(02)02138-X} {\bibfield  {journal} {\bibinfo
  {journal} {Phys.\ Lett.\ B}\ }\textbf {\bibinfo {volume} {540}},\ \bibinfo
  {pages} {14} (\bibinfo {year} {2002})},\ \Eprint
  {http://arXiv.org/abs/hep-ph/0204286} {arXiv:hep-ph/0204286} \BibitemShut
  {NoStop}%
\bibitem [{\citenamefont {Picoreti}\ \emph {et~al.}(2016)\citenamefont
  {Picoreti}, \citenamefont {Guzzo}, \citenamefont {de~Holanda},\ and\
  \citenamefont {Peres}}]{Picoreti:2015ika}%
  \BibitemOpen
  \bibfield  {author} {\bibinfo {author} {\bibfnamefont {R.}~\bibnamefont
  {Picoreti}}, \bibinfo {author} {\bibfnamefont {M.~M.}\ \bibnamefont {Guzzo}},
  \bibinfo {author} {\bibfnamefont {P.}~\bibnamefont {de~Holanda}}, \ and\
  \bibinfo {author} {\bibfnamefont {O.}~\bibnamefont {Peres}},\ }\href
  {\doibase 10.1016/j.physletb.2016.08.007} {\bibfield  {journal} {\bibinfo
  {journal} {Phys.\ Lett.\ B}\ }\textbf {\bibinfo {volume} {761}},\ \bibinfo
  {pages} {70} (\bibinfo {year} {2016})},\ \Eprint
  {http://arxiv.org/abs/1506.08158} {arXiv:1506.08158 [hep-ph]} \BibitemShut
  {NoStop}%
\bibitem [{\citenamefont {Aharmim}\ \emph {et~al.}(2019)\citenamefont {Aharmim}
  \emph {et~al.}}]{Aharmim:2018fme}%
  \BibitemOpen
  \bibfield  {author} {\bibinfo {author} {\bibfnamefont {B.}~\bibnamefont
  {Aharmim}} \emph {et~al.} (\bibinfo {collaboration} {SNO}),\ }\href {\doibase
  10.1103/PhysRevD.99.032013} {\bibfield  {journal} {\bibinfo  {journal}
  {Phys.\ Rev.\ D}\ }\textbf {\bibinfo {volume} {99}},\ \bibinfo {pages}
  {032013} (\bibinfo {year} {2019})},\ \Eprint
  {http://arxiv.org/abs/1812.01088} {arXiv:1812.01088 [hep-ex]} \BibitemShut
  {NoStop}%
\bibitem [{\citenamefont {Gonzalez-Garcia}\ and\ \citenamefont
  {Maltoni}(2008)}]{GonzalezGarcia:2008ru}%
  \BibitemOpen
  \bibfield  {author} {\bibinfo {author} {\bibfnamefont {M.~C.}\ \bibnamefont
  {Gonzalez-Garcia}}\ and\ \bibinfo {author} {\bibfnamefont {M.}~\bibnamefont
  {Maltoni}},\ }\href {\doibase 10.1016/j.physletb.2008.04.041} {\bibfield
  {journal} {\bibinfo  {journal} {Phys.~Lett.~B}\ }\textbf {\bibinfo {volume}
  {663}},\ \bibinfo {pages} {405} (\bibinfo {year} {2008})},\ \Eprint
  {http://arxiv.org/abs/0802.3699} {arXiv:0802.3699 [hep-ph]} \BibitemShut
  {NoStop}%
\bibitem [{\citenamefont {Gomes}\ \emph {et~al.}(2015)\citenamefont {Gomes},
  \citenamefont {Gomes},\ and\ \citenamefont {Peres}}]{Gomes:2014yua}%
  \BibitemOpen
  \bibfield  {author} {\bibinfo {author} {\bibfnamefont {R.~A.}\ \bibnamefont
  {Gomes}}, \bibinfo {author} {\bibfnamefont {A.~L.~G.}\ \bibnamefont {Gomes}},
  \ and\ \bibinfo {author} {\bibfnamefont {O.}~\bibnamefont {Peres}},\ }\href
  {\doibase 10.1016/j.physletb.2014.12.014} {\bibfield  {journal} {\bibinfo
  {journal} {Phys.~Lett.~B}\ }\textbf {\bibinfo {volume} {740}},\ \bibinfo
  {pages} {345} (\bibinfo {year} {2015})},\ \Eprint
  {http://arxiv.org/abs/1407.5640} {arXiv:1407.5640 [hep-ph]} \BibitemShut
  {NoStop}%
\bibitem [{\citenamefont {Porto-Silva}\ \emph {et~al.}(2020)\citenamefont
  {Porto-Silva}, \citenamefont {Prakash}, \citenamefont {Peres}, \citenamefont
  {Nunokawa},\ and\ \citenamefont {Minakata}}]{Porto-Silva:2020gma}%
  \BibitemOpen
  \bibfield  {author} {\bibinfo {author} {\bibfnamefont {Y.~P.}\ \bibnamefont
  {Porto-Silva}}, \bibinfo {author} {\bibfnamefont {S.}~\bibnamefont
  {Prakash}}, \bibinfo {author} {\bibfnamefont {O.}~\bibnamefont {Peres}},
  \bibinfo {author} {\bibfnamefont {H.}~\bibnamefont {Nunokawa}}, \ and\
  \bibinfo {author} {\bibfnamefont {H.}~\bibnamefont {Minakata}},\ }\href@noop
  {} {\  (\bibinfo {year} {2020})},\ \Eprint {http://arxiv.org/abs/2002.12134}
  {arXiv:2002.12134 [hep-ph]} \BibitemShut {NoStop}%
\bibitem [{\citenamefont {Gago}\ \emph {et~al.}(2017)\citenamefont {Gago},
  \citenamefont {Gomes}, \citenamefont {Gomes}, \citenamefont {Jones-Perez},\
  and\ \citenamefont {Peres}}]{Gago:2017zzy}%
  \BibitemOpen
  \bibfield  {author} {\bibinfo {author} {\bibfnamefont {A.~M.}\ \bibnamefont
  {Gago}}, \bibinfo {author} {\bibfnamefont {R.~A.}\ \bibnamefont {Gomes}},
  \bibinfo {author} {\bibfnamefont {A.~L.~G.}\ \bibnamefont {Gomes}}, \bibinfo
  {author} {\bibfnamefont {J.}~\bibnamefont {Jones-Perez}}, \ and\ \bibinfo
  {author} {\bibfnamefont {O.~L.~G.}\ \bibnamefont {Peres}},\ }\href {\doibase
  10.1007/JHEP11(2017)022} {\bibfield  {journal} {\bibinfo  {journal} {JHEP}\
  }\textbf {\bibinfo {volume} {11}},\ \bibinfo {pages} {022} (\bibinfo {year}
  {2017})},\ \Eprint {http://arxiv.org/abs/1705.03074} {arXiv:1705.03074
  [hep-ph]} \BibitemShut {NoStop}%
\bibitem [{\citenamefont {Coloma}\ and\ \citenamefont
  {Peres}(2017)}]{Coloma:2017zpg}%
  \BibitemOpen
  \bibfield  {author} {\bibinfo {author} {\bibfnamefont {P.}~\bibnamefont
  {Coloma}}\ and\ \bibinfo {author} {\bibfnamefont {O.}~\bibnamefont {Peres}},\
  }\href@noop {} {\  (\bibinfo {year} {2017})},\ \Eprint
  {http://arxiv.org/abs/1705.03599} {arXiv:1705.03599 [hep-ph]} \BibitemShut
  {NoStop}%
\bibitem [{\citenamefont {Lu}()}]{Talk_Lu_UHECR_2018}%
  \BibitemOpen
  \bibfield  {author} {\bibinfo {author} {\bibfnamefont {L.}~\bibnamefont
  {Lu}},\ }\href@noop {} {\ }\bibinfo {note} {Presented at UHECR 2018, October
  10, 2018, Paris, {\tt https://indico.in2p3.fr/event/17063/}}\BibitemShut
  {NoStop}%
\bibitem [{\citenamefont {De~Salas}\ \emph {et~al.}(2018)\citenamefont
  {De~Salas}, \citenamefont {Gariazzo}, \citenamefont {Mena}, \citenamefont
  {Ternes},\ and\ \citenamefont {T\'ortola}}]{deSalas:2018bym}%
  \BibitemOpen
  \bibfield  {author} {\bibinfo {author} {\bibfnamefont {P.~F.}\ \bibnamefont
  {De~Salas}}, \bibinfo {author} {\bibfnamefont {S.}~\bibnamefont {Gariazzo}},
  \bibinfo {author} {\bibfnamefont {O.}~\bibnamefont {Mena}}, \bibinfo {author}
  {\bibfnamefont {C.~A.}\ \bibnamefont {Ternes}}, \ and\ \bibinfo {author}
  {\bibfnamefont {M.}~\bibnamefont {T\'ortola}},\ }\href {\doibase
  10.3389/fspas.2018.00036} {\bibfield  {journal} {\bibinfo  {journal} {Front.\
  Astron.\ Space Sci.}\ }\textbf {\bibinfo {volume} {5}},\ \bibinfo {pages}
  {36} (\bibinfo {year} {2018})},\ \Eprint {http://arxiv.org/abs/1806.11051}
  {arXiv:1806.11051 [hep-ph]} \BibitemShut {NoStop}%
\bibitem [{\citenamefont {NuFit}()}]{NuFit_4.1}%
  \BibitemOpen
  \bibfield  {author} {\bibinfo {author} {\bibnamefont {NuFit}},\ }\href@noop
  {} {\enquote {\bibinfo {title} {{Three-neutrino fit based on data available
  in July 2019}},}\ }\bibinfo {note} {{\tt http://www.nu-fit.org/}}\BibitemShut
  {NoStop}%
\bibitem [{\citenamefont {Roy~C.}\ and\ \citenamefont
  {Choubey}(2018)}]{Choudhury:2018byy}%
  \BibitemOpen
  \bibfield  {author} {\bibinfo {author} {\bibfnamefont {S.}~\bibnamefont
  {Roy~C.}}\ and\ \bibinfo {author} {\bibfnamefont {S.}~\bibnamefont
  {Choubey}},\ }\href {\doibase 10.1088/1475-7516/2018/09/017} {\bibfield
  {journal} {\bibinfo  {journal} {JCAP}\ }\textbf {\bibinfo {volume} {09}},\
  \bibinfo {pages} {017} (\bibinfo {year} {2018})},\ \Eprint
  {http://arxiv.org/abs/1806.10832} {arXiv:1806.10832 [astro-ph.CO]}
  \BibitemShut {NoStop}%
\bibitem [{\citenamefont {Glashow}(1960)}]{Glashow:1960zz}%
  \BibitemOpen
  \bibfield  {author} {\bibinfo {author} {\bibfnamefont {S.~L.}\ \bibnamefont
  {Glashow}},\ }\href {\doibase 10.1103/PhysRev.118.316} {\bibfield  {journal}
  {\bibinfo  {journal} {Phys.\ Rev.}\ }\textbf {\bibinfo {volume} {118}},\
  \bibinfo {pages} {316} (\bibinfo {year} {1960})}\BibitemShut {NoStop}%
\bibitem [{\citenamefont {Bhattacharya}\ \emph {et~al.}(2011)\citenamefont
  {Bhattacharya}, \citenamefont {Gandhi}, \citenamefont {Rodejohann},\ and\
  \citenamefont {Watanabe}}]{Bhattacharya:2011qu}%
  \BibitemOpen
  \bibfield  {author} {\bibinfo {author} {\bibfnamefont {A.}~\bibnamefont
  {Bhattacharya}}, \bibinfo {author} {\bibfnamefont {R.}~\bibnamefont
  {Gandhi}}, \bibinfo {author} {\bibfnamefont {W.}~\bibnamefont {Rodejohann}},
  \ and\ \bibinfo {author} {\bibfnamefont {A.}~\bibnamefont {Watanabe}},\
  }\href {\doibase 10.1088/1475-7516/2011/10/017} {\bibfield  {journal}
  {\bibinfo  {journal} {JCAP}\ }\textbf {\bibinfo {volume} {1110}},\ \bibinfo
  {pages} {017} (\bibinfo {year} {2011})},\ \Eprint
  {http://arxiv.org/abs/1108.3163} {arXiv:1108.3163 [astro-ph.HE]} \BibitemShut
  {NoStop}%
\bibitem [{\citenamefont {Barger}\ \emph {et~al.}(2013)\citenamefont {Barger},
  \citenamefont {Learned},\ and\ \citenamefont {Pakvasa}}]{Barger:2012mz}%
  \BibitemOpen
  \bibfield  {author} {\bibinfo {author} {\bibfnamefont {V.}~\bibnamefont
  {Barger}}, \bibinfo {author} {\bibfnamefont {J.}~\bibnamefont {Learned}}, \
  and\ \bibinfo {author} {\bibfnamefont {S.}~\bibnamefont {Pakvasa}},\ }\href
  {\doibase 10.1103/PhysRevD.87.037302} {\bibfield  {journal} {\bibinfo
  {journal} {Phys.\ Rev.\ D}\ }\textbf {\bibinfo {volume} {87}},\ \bibinfo
  {pages} {037302} (\bibinfo {year} {2013})},\ \Eprint
  {http://arxiv.org/abs/1207.4571} {arXiv:1207.4571 [astro-ph.HE]} \BibitemShut
  {NoStop}%
\bibitem [{\citenamefont {Bhattacharya}\ \emph {et~al.}(2012)\citenamefont
  {Bhattacharya}, \citenamefont {Gandhi}, \citenamefont {Rodejohann},\ and\
  \citenamefont {Watanabe}}]{Bhattacharya:2012fh}%
  \BibitemOpen
  \bibfield  {author} {\bibinfo {author} {\bibfnamefont {A.}~\bibnamefont
  {Bhattacharya}}, \bibinfo {author} {\bibfnamefont {R.}~\bibnamefont
  {Gandhi}}, \bibinfo {author} {\bibfnamefont {W.}~\bibnamefont {Rodejohann}},
  \ and\ \bibinfo {author} {\bibfnamefont {A.}~\bibnamefont {Watanabe}},\
  }\href@noop {} {\  (\bibinfo {year} {2012})},\ \Eprint
  {http://arxiv.org/abs/1209.2422} {arXiv:1209.2422 [hep-ph]} \BibitemShut
  {NoStop}%
\bibitem [{\citenamefont {Barger}\ \emph {et~al.}(2014)\citenamefont {Barger},
  \citenamefont {Fu}, \citenamefont {Learned}, \citenamefont {Marfatia},
  \citenamefont {Pakvasa},\ and\ \citenamefont {Weiler}}]{Barger:2014iua}%
  \BibitemOpen
  \bibfield  {author} {\bibinfo {author} {\bibfnamefont {V.}~\bibnamefont
  {Barger}}, \bibinfo {author} {\bibfnamefont {L.}~\bibnamefont {Fu}}, \bibinfo
  {author} {\bibfnamefont {J.~G.}\ \bibnamefont {Learned}}, \bibinfo {author}
  {\bibfnamefont {D.}~\bibnamefont {Marfatia}}, \bibinfo {author}
  {\bibfnamefont {S.}~\bibnamefont {Pakvasa}}, \ and\ \bibinfo {author}
  {\bibfnamefont {T.~J.}\ \bibnamefont {Weiler}},\ }\href {\doibase
  10.1103/PhysRevD.90.121301} {\bibfield  {journal} {\bibinfo  {journal}
  {Phys.\ Rev.\ D}\ }\textbf {\bibinfo {volume} {90}},\ \bibinfo {pages}
  {121301} (\bibinfo {year} {2014})},\ \Eprint {http://arxiv.org/abs/1407.3255}
  {arXiv:1407.3255 [astro-ph.HE]} \BibitemShut {NoStop}%
\bibitem [{\citenamefont {Huang}\ and\ \citenamefont
  {Liu}(2020)}]{Huang:2019hgs}%
  \BibitemOpen
  \bibfield  {author} {\bibinfo {author} {\bibfnamefont {G.-y.}\ \bibnamefont
  {Huang}}\ and\ \bibinfo {author} {\bibfnamefont {Q.}~\bibnamefont {Liu}},\
  }\href {\doibase 10.1088/1475-7516/2020/03/005} {\bibfield  {journal}
  {\bibinfo  {journal} {JCAP}\ }\textbf {\bibinfo {volume} {03}},\ \bibinfo
  {pages} {005} (\bibinfo {year} {2020})},\ \Eprint
  {http://arxiv.org/abs/1912.02976} {arXiv:1912.02976 [hep-ph]} \BibitemShut
  {NoStop}%
\bibitem [{\citenamefont {Capozzi}\ \emph {et~al.}(2017)\citenamefont
  {Capozzi}, \citenamefont {Di~Valentino}, \citenamefont {Lisi}, \citenamefont
  {Marrone}, \citenamefont {Melchiorri},\ and\ \citenamefont
  {Palazzo}}]{Capozzi:2017ipn}%
  \BibitemOpen
  \bibfield  {author} {\bibinfo {author} {\bibfnamefont {F.}~\bibnamefont
  {Capozzi}}, \bibinfo {author} {\bibfnamefont {E.}~\bibnamefont
  {Di~Valentino}}, \bibinfo {author} {\bibfnamefont {E.}~\bibnamefont {Lisi}},
  \bibinfo {author} {\bibfnamefont {A.}~\bibnamefont {Marrone}}, \bibinfo
  {author} {\bibfnamefont {A.}~\bibnamefont {Melchiorri}}, \ and\ \bibinfo
  {author} {\bibfnamefont {A.}~\bibnamefont {Palazzo}},\ }\href {\doibase
  10.1103/PhysRevD.95.096014} {\bibfield  {journal} {\bibinfo  {journal}
  {Phys.\ Rev.\ D}\ }\textbf {\bibinfo {volume} {95}},\ \bibinfo {pages}
  {096014} (\bibinfo {year} {2017})},\ \Eprint
  {http://arxiv.org/abs/1703.04471} {arXiv:1703.04471 [hep-ph]} \BibitemShut
  {NoStop}%
\bibitem [{\citenamefont {Maki}\ \emph {et~al.}(1962)\citenamefont {Maki},
  \citenamefont {Nakagawa},\ and\ \citenamefont {Sakata}}]{Maki:1962mu}%
  \BibitemOpen
  \bibfield  {author} {\bibinfo {author} {\bibfnamefont {Z.}~\bibnamefont
  {Maki}}, \bibinfo {author} {\bibfnamefont {M.}~\bibnamefont {Nakagawa}}, \
  and\ \bibinfo {author} {\bibfnamefont {S.}~\bibnamefont {Sakata}},\ }\href
  {\doibase 10.1143/PTP.28.870} {\bibfield  {journal} {\bibinfo  {journal}
  {Prog.\ Theor.\ Phys.}\ }\textbf {\bibinfo {volume} {28}},\ \bibinfo {pages}
  {870} (\bibinfo {year} {1962})}\BibitemShut {NoStop}%
\bibitem [{\citenamefont {Pontecorvo}(1968)}]{Pontecorvo:1967fh}%
  \BibitemOpen
  \bibfield  {author} {\bibinfo {author} {\bibfnamefont {B.}~\bibnamefont
  {Pontecorvo}},\ }\href@noop {} {\bibfield  {journal} {\bibinfo  {journal}
  {Sov.\ Phys.\ JETP}\ }\textbf {\bibinfo {volume} {26}},\ \bibinfo {pages}
  {984} (\bibinfo {year} {1968})}\BibitemShut {NoStop}%
\bibitem [{\citenamefont {Tanabashi}\ \emph {et~al.}(2018)\citenamefont
  {Tanabashi} \emph {et~al.}}]{Tanabashi:2018oca}%
  \BibitemOpen
  \bibfield  {author} {\bibinfo {author} {\bibfnamefont {M.}~\bibnamefont
  {Tanabashi}} \emph {et~al.} (\bibinfo {collaboration} {Particle Data
  Group}),\ }\href {\doibase 10.1103/PhysRevD.98.030001} {\bibfield  {journal}
  {\bibinfo  {journal} {Phys.\ Rev.\ D}\ }\textbf {\bibinfo {volume} {98}},\
  \bibinfo {pages} {030001} (\bibinfo {year} {2018})}\BibitemShut {NoStop}%
\bibitem [{\citenamefont {Capozzi}\ \emph {et~al.}(2018)\citenamefont
  {Capozzi}, \citenamefont {Lisi}, \citenamefont {Marrone},\ and\ \citenamefont
  {Palazzo}}]{Capozzi:2018ubv}%
  \BibitemOpen
  \bibfield  {author} {\bibinfo {author} {\bibfnamefont {F.}~\bibnamefont
  {Capozzi}}, \bibinfo {author} {\bibfnamefont {E.}~\bibnamefont {Lisi}},
  \bibinfo {author} {\bibfnamefont {A.}~\bibnamefont {Marrone}}, \ and\
  \bibinfo {author} {\bibfnamefont {A.}~\bibnamefont {Palazzo}},\ }\href
  {\doibase 10.1016/j.ppnp.2018.05.005} {\bibfield  {journal} {\bibinfo
  {journal} {Prog.\ Part.\ Nucl.\ Phys.}\ }\textbf {\bibinfo {volume} {102}},\
  \bibinfo {pages} {48} (\bibinfo {year} {2018})},\ \Eprint
  {http://arxiv.org/abs/1804.09678} {arXiv:1804.09678 [hep-ph]} \BibitemShut
  {NoStop}%
\bibitem [{\citenamefont {Esteban}\ \emph {et~al.}(2019)\citenamefont
  {Esteban}, \citenamefont {Gonzalez-Garcia}, \citenamefont
  {Hernandez-Cabezudo}, \citenamefont {Maltoni},\ and\ \citenamefont
  {Schwetz}}]{Esteban:2018azc}%
  \BibitemOpen
  \bibfield  {author} {\bibinfo {author} {\bibfnamefont {I.}~\bibnamefont
  {Esteban}}, \bibinfo {author} {\bibfnamefont {M.~C.}\ \bibnamefont
  {Gonzalez-Garcia}}, \bibinfo {author} {\bibfnamefont {A.}~\bibnamefont
  {Hernandez-Cabezudo}}, \bibinfo {author} {\bibfnamefont {M.}~\bibnamefont
  {Maltoni}}, \ and\ \bibinfo {author} {\bibfnamefont {T.}~\bibnamefont
  {Schwetz}},\ }\href {\doibase 10.1007/JHEP01(2019)106} {\bibfield  {journal}
  {\bibinfo  {journal} {JHEP}\ }\textbf {\bibinfo {volume} {01}},\ \bibinfo
  {pages} {106} (\bibinfo {year} {2019})},\ \Eprint
  {http://arxiv.org/abs/1811.05487} {arXiv:1811.05487 [hep-ph]} \BibitemShut
  {NoStop}%
\bibitem [{\citenamefont {Pakvasa}(2008)}]{Pakvasa:2008nx}%
  \BibitemOpen
  \bibfield  {author} {\bibinfo {author} {\bibfnamefont {S.}~\bibnamefont
  {Pakvasa}},\ }\href {\doibase 10.1142/S0217732308027680} {\bibfield
  {journal} {\bibinfo  {journal} {Mod.\ Phys.\ Lett.\ A}\ }\textbf {\bibinfo
  {volume} {23}},\ \bibinfo {pages} {1313} (\bibinfo {year} {2008})},\ \Eprint
  {http://arxiv.org/abs/0803.1701} {arXiv:0803.1701 [hep-ph]} \BibitemShut
  {NoStop}%
\bibitem [{\citenamefont {Margolis}\ \emph {et~al.}(1978)\citenamefont
  {Margolis}, \citenamefont {Schramm},\ and\ \citenamefont
  {Silberberg}}]{Margolis:1977wt}%
  \BibitemOpen
  \bibfield  {author} {\bibinfo {author} {\bibfnamefont {S.~H.}\ \bibnamefont
  {Margolis}}, \bibinfo {author} {\bibfnamefont {D.~N.}\ \bibnamefont
  {Schramm}}, \ and\ \bibinfo {author} {\bibfnamefont {R.}~\bibnamefont
  {Silberberg}},\ }\href {\doibase 10.1086/156104} {\bibfield  {journal}
  {\bibinfo  {journal} {Astrophys.\ J.}\ }\textbf {\bibinfo {volume} {221}},\
  \bibinfo {pages} {990} (\bibinfo {year} {1978})}\BibitemShut {NoStop}%
\bibitem [{\citenamefont {Stecker}(1979)}]{Stecker:1978ah}%
  \BibitemOpen
  \bibfield  {author} {\bibinfo {author} {\bibfnamefont {F.~W.}\ \bibnamefont
  {Stecker}},\ }\href {\doibase 10.1086/156919} {\bibfield  {journal} {\bibinfo
   {journal} {Astrophys. J.}\ }\textbf {\bibinfo {volume} {228}},\ \bibinfo
  {pages} {919} (\bibinfo {year} {1979})}\BibitemShut {NoStop}%
\bibitem [{\citenamefont {Kelner}\ \emph {et~al.}(2006)\citenamefont {Kelner},
  \citenamefont {Aharonian},\ and\ \citenamefont {Bugayov}}]{Kelner:2006tc}%
  \BibitemOpen
  \bibfield  {author} {\bibinfo {author} {\bibfnamefont {S.~R.}\ \bibnamefont
  {Kelner}}, \bibinfo {author} {\bibfnamefont {F.~A.}\ \bibnamefont
  {Aharonian}}, \ and\ \bibinfo {author} {\bibfnamefont {V.~V.}\ \bibnamefont
  {Bugayov}},\ }\href {\doibase 10.1103/PhysRevD.74.034018,
  10.1103/PhysRevD.79.039901} {\bibfield  {journal} {\bibinfo  {journal}
  {Phys.\ Rev.\ D}\ }\textbf {\bibinfo {volume} {74}},\ \bibinfo {pages}
  {034018} (\bibinfo {year} {2006})},\ \bibinfo {note} {[Erratum: Phys.\ Rev.\
  D 79, 039901 (2009)]},\ \Eprint {http://arxiv.org/abs/astro-ph/0606058}
  {arXiv:astro-ph/0606058 [astro-ph]} \BibitemShut {NoStop}%
\bibitem [{\citenamefont {M{\"u}cke}\ \emph {et~al.}(2000)\citenamefont
  {M{\"u}cke}, \citenamefont {Engel}, \citenamefont {Rachen}, \citenamefont
  {Protheroe},\ and\ \citenamefont {Stanev}}]{Mucke:1999yb}%
  \BibitemOpen
  \bibfield  {author} {\bibinfo {author} {\bibfnamefont {A.}~\bibnamefont
  {M{\"u}cke}}, \bibinfo {author} {\bibfnamefont {R.}~\bibnamefont {Engel}},
  \bibinfo {author} {\bibfnamefont {J.}~\bibnamefont {Rachen}}, \bibinfo
  {author} {\bibfnamefont {R.}~\bibnamefont {Protheroe}}, \ and\ \bibinfo
  {author} {\bibfnamefont {T.}~\bibnamefont {Stanev}},\ }\href {\doibase
  10.1016/S0010-4655(99)00446-4} {\bibfield  {journal} {\bibinfo  {journal}
  {Comput.\ Phys.\ Commun.}\ }\textbf {\bibinfo {volume} {124}},\ \bibinfo
  {pages} {290} (\bibinfo {year} {2000})},\ \Eprint
  {http://arxiv.org/abs/astro-ph/9903478} {arXiv:astro-ph/9903478 [astro-ph]}
  \BibitemShut {NoStop}%
\bibitem [{\citenamefont {H{\"u}mmer}\ \emph {et~al.}(2010)\citenamefont
  {H{\"u}mmer}, \citenamefont {R{\"u}ger}, \citenamefont {Spanier},\ and\
  \citenamefont {Winter}}]{Hummer:2010vx}%
  \BibitemOpen
  \bibfield  {author} {\bibinfo {author} {\bibfnamefont {S.}~\bibnamefont
  {H{\"u}mmer}}, \bibinfo {author} {\bibfnamefont {M.}~\bibnamefont
  {R{\"u}ger}}, \bibinfo {author} {\bibfnamefont {F.}~\bibnamefont {Spanier}},
  \ and\ \bibinfo {author} {\bibfnamefont {W.}~\bibnamefont {Winter}},\ }\href
  {\doibase 10.1088/0004-637X/721/1/630} {\bibfield  {journal} {\bibinfo
  {journal} {Astrophys.\ J.}\ }\textbf {\bibinfo {volume} {721}},\ \bibinfo
  {pages} {630} (\bibinfo {year} {2010})},\ \Eprint
  {http://arxiv.org/abs/1002.1310} {arXiv:1002.1310 [astro-ph.HE]} \BibitemShut
  {NoStop}%
\bibitem [{\citenamefont {Bustamante}\ and\ \citenamefont
  {Ahlers}(2019)}]{Bustamante:2019sdb}%
  \BibitemOpen
  \bibfield  {author} {\bibinfo {author} {\bibfnamefont {M.}~\bibnamefont
  {Bustamante}}\ and\ \bibinfo {author} {\bibfnamefont {M.}~\bibnamefont
  {Ahlers}},\ }\href {\doibase 10.1103/PhysRevLett.122.241101} {\bibfield
  {journal} {\bibinfo  {journal} {Phys.\ Rev.\ Lett.}\ }\textbf {\bibinfo
  {volume} {122}},\ \bibinfo {pages} {241101} (\bibinfo {year} {2019})},\
  \Eprint {http://arxiv.org/abs/1901.10087} {arXiv:1901.10087 [astro-ph.HE]}
  \BibitemShut {NoStop}%
\bibitem [{\citenamefont {Schechter}\ and\ \citenamefont
  {Valle}(1982)}]{Schechter:1981cv}%
  \BibitemOpen
  \bibfield  {author} {\bibinfo {author} {\bibfnamefont {J.}~\bibnamefont
  {Schechter}}\ and\ \bibinfo {author} {\bibfnamefont {J.~W.~F.}\ \bibnamefont
  {Valle}},\ }\href {\doibase 10.1103/PhysRevD.25.774} {\bibfield  {journal}
  {\bibinfo  {journal} {Phys.\ Rev.\ D}\ }\textbf {\bibinfo {volume} {25}},\
  \bibinfo {pages} {774} (\bibinfo {year} {1982})}\BibitemShut {NoStop}%
\bibitem [{\citenamefont {Berryman}\ \emph {et~al.}(2018)\citenamefont
  {Berryman}, \citenamefont {De~Gouvêa}, \citenamefont {Kelly},\ and\
  \citenamefont {Zhang}}]{Berryman:2018ogk}%
  \BibitemOpen
  \bibfield  {author} {\bibinfo {author} {\bibfnamefont {J.~M.}\ \bibnamefont
  {Berryman}}, \bibinfo {author} {\bibfnamefont {A.}~\bibnamefont
  {De~Gouvêa}}, \bibinfo {author} {\bibfnamefont {K.~J.}\ \bibnamefont
  {Kelly}}, \ and\ \bibinfo {author} {\bibfnamefont {Y.}~\bibnamefont
  {Zhang}},\ }\href {\doibase 10.1103/PhysRevD.97.075030} {\bibfield  {journal}
  {\bibinfo  {journal} {Phys.\ Rev.\ D}\ }\textbf {\bibinfo {volume} {97}},\
  \bibinfo {pages} {075030} (\bibinfo {year} {2018})},\ \Eprint
  {http://arxiv.org/abs/1802.00009} {arXiv:1802.00009 [hep-ph]} \BibitemShut
  {NoStop}%
\bibitem [{\citenamefont {de~Gouvêa}\ \emph {et~al.}(2020)\citenamefont
  {de~Gouvêa}, \citenamefont {Martinez-Soler},\ and\ \citenamefont
  {Sen}}]{deGouvea:2019goq}%
  \BibitemOpen
  \bibfield  {author} {\bibinfo {author} {\bibfnamefont {A.}~\bibnamefont
  {de~Gouvêa}}, \bibinfo {author} {\bibfnamefont {I.}~\bibnamefont
  {Martinez-Soler}}, \ and\ \bibinfo {author} {\bibfnamefont {M.}~\bibnamefont
  {Sen}},\ }\href {\doibase 10.1103/PhysRevD.101.043013} {\bibfield  {journal}
  {\bibinfo  {journal} {Phys.\ Rev.\ D}\ }\textbf {\bibinfo {volume} {101}},\
  \bibinfo {pages} {043013} (\bibinfo {year} {2020})},\ \Eprint
  {http://arxiv.org/abs/1910.01127} {arXiv:1910.01127 [hep-ph]} \BibitemShut
  {NoStop}%
\bibitem [{\citenamefont {Capozzi}\ \emph {et~al.}(2020)\citenamefont
  {Capozzi}, \citenamefont {Di~Valentino}, \citenamefont {Lisi}, \citenamefont
  {Marrone}, \citenamefont {Melchiorri},\ and\ \citenamefont
  {Palazzo}}]{Capozzi:2020qhw}%
  \BibitemOpen
  \bibfield  {author} {\bibinfo {author} {\bibfnamefont {F.}~\bibnamefont
  {Capozzi}}, \bibinfo {author} {\bibfnamefont {E.}~\bibnamefont
  {Di~Valentino}}, \bibinfo {author} {\bibfnamefont {E.}~\bibnamefont {Lisi}},
  \bibinfo {author} {\bibfnamefont {A.}~\bibnamefont {Marrone}}, \bibinfo
  {author} {\bibfnamefont {A.}~\bibnamefont {Melchiorri}}, \ and\ \bibinfo
  {author} {\bibfnamefont {A.}~\bibnamefont {Palazzo}},\ }\href@noop {} {\
  (\bibinfo {year} {2020})},\ \Eprint {http://arxiv.org/abs/2003.08511}
  {arXiv:2003.08511 [hep-ph]} \BibitemShut {NoStop}%
\bibitem [{\citenamefont {Ahlers}\ and\ \citenamefont
  {Murase}(2014)}]{Ahlers:2013xia}%
  \BibitemOpen
  \bibfield  {author} {\bibinfo {author} {\bibfnamefont {M.}~\bibnamefont
  {Ahlers}}\ and\ \bibinfo {author} {\bibfnamefont {K.}~\bibnamefont
  {Murase}},\ }\href {\doibase 10.1103/PhysRevD.90.023010} {\bibfield
  {journal} {\bibinfo  {journal} {Phys.\ Rev.\ D}\ }\textbf {\bibinfo {volume}
  {90}},\ \bibinfo {pages} {023010} (\bibinfo {year} {2014})},\ \Eprint
  {http://arxiv.org/abs/1309.4077} {arXiv:1309.4077 [astro-ph.HE]} \BibitemShut
  {NoStop}%
\bibitem [{\citenamefont {Ahlers}\ \emph {et~al.}(2016)\citenamefont {Ahlers},
  \citenamefont {Bai}, \citenamefont {Barger},\ and\ \citenamefont
  {Lu}}]{Ahlers:2015moa}%
  \BibitemOpen
  \bibfield  {author} {\bibinfo {author} {\bibfnamefont {M.}~\bibnamefont
  {Ahlers}}, \bibinfo {author} {\bibfnamefont {Y.}~\bibnamefont {Bai}},
  \bibinfo {author} {\bibfnamefont {V.}~\bibnamefont {Barger}}, \ and\ \bibinfo
  {author} {\bibfnamefont {R.}~\bibnamefont {Lu}},\ }\href {\doibase
  10.1103/PhysRevD.93.013009} {\bibfield  {journal} {\bibinfo  {journal}
  {Phys.\ Rev.\ D}\ }\textbf {\bibinfo {volume} {93}},\ \bibinfo {pages}
  {013009} (\bibinfo {year} {2016})},\ \Eprint
  {http://arxiv.org/abs/1505.03156} {arXiv:1505.03156 [hep-ph]} \BibitemShut
  {NoStop}%
\bibitem [{\citenamefont {Murase}\ \emph {et~al.}(2016)\citenamefont {Murase},
  \citenamefont {Guetta},\ and\ \citenamefont {Ahlers}}]{Murase:2015xka}%
  \BibitemOpen
  \bibfield  {author} {\bibinfo {author} {\bibfnamefont {K.}~\bibnamefont
  {Murase}}, \bibinfo {author} {\bibfnamefont {D.}~\bibnamefont {Guetta}}, \
  and\ \bibinfo {author} {\bibfnamefont {M.}~\bibnamefont {Ahlers}},\ }\href
  {\doibase 10.1103/PhysRevLett.116.071101} {\bibfield  {journal} {\bibinfo
  {journal} {Phys.\ Rev.\ Lett.}\ }\textbf {\bibinfo {volume} {116}},\ \bibinfo
  {pages} {071101} (\bibinfo {year} {2016})},\ \Eprint
  {http://arxiv.org/abs/1509.00805} {arXiv:1509.00805 [astro-ph.HE]}
  \BibitemShut {NoStop}%
\bibitem [{\citenamefont {Denton}\ \emph {et~al.}(2017)\citenamefont {Denton},
  \citenamefont {Marfatia},\ and\ \citenamefont {Weiler}}]{Denton:2017csz}%
  \BibitemOpen
  \bibfield  {author} {\bibinfo {author} {\bibfnamefont {P.~B.}\ \bibnamefont
  {Denton}}, \bibinfo {author} {\bibfnamefont {D.}~\bibnamefont {Marfatia}}, \
  and\ \bibinfo {author} {\bibfnamefont {T.~J.}\ \bibnamefont {Weiler}},\
  }\href {\doibase 10.1088/1475-7516/2017/08/033} {\bibfield  {journal}
  {\bibinfo  {journal} {JCAP}\ }\textbf {\bibinfo {volume} {1708}},\ \bibinfo
  {pages} {033} (\bibinfo {year} {2017})},\ \Eprint
  {http://arxiv.org/abs/1703.09721} {arXiv:1703.09721 [astro-ph.HE]}
  \BibitemShut {NoStop}%
\bibitem [{\citenamefont {Aartsen}\ \emph {et~al.}(2017)\citenamefont {Aartsen}
  \emph {et~al.}}]{Aartsen:2017ujz}%
  \BibitemOpen
  \bibfield  {author} {\bibinfo {author} {\bibfnamefont {M.~G.}\ \bibnamefont
  {Aartsen}} \emph {et~al.} (\bibinfo {collaboration} {IceCube}),\ }\href
  {\doibase 10.3847/1538-4357/aa8dfb} {\bibfield  {journal} {\bibinfo
  {journal} {Astrophys.\ J.}\ }\textbf {\bibinfo {volume} {849}},\ \bibinfo
  {pages} {67} (\bibinfo {year} {2017})},\ \Eprint
  {http://arxiv.org/abs/1707.03416} {arXiv:1707.03416 [astro-ph.HE]}
  \BibitemShut {NoStop}%
\bibitem [{\citenamefont {Ahlers}\ and\ \citenamefont
  {Halzen}(2018)}]{Ahlers:2018fkn}%
  \BibitemOpen
  \bibfield  {author} {\bibinfo {author} {\bibfnamefont {M.}~\bibnamefont
  {Ahlers}}\ and\ \bibinfo {author} {\bibfnamefont {F.}~\bibnamefont
  {Halzen}},\ }\href {\doibase 10.1016/j.ppnp.2018.05.001} {\bibfield
  {journal} {\bibinfo  {journal} {Prog.\ Part.\ Nucl.\ Phys.}\ }\textbf
  {\bibinfo {volume} {102}},\ \bibinfo {pages} {73} (\bibinfo {year} {2018})},\
  \Eprint {http://arxiv.org/abs/1805.11112} {arXiv:1805.11112 [astro-ph.HE]}
  \BibitemShut {NoStop}%
\bibitem [{\citenamefont {Yuksel}\ \emph {et~al.}(2008)\citenamefont {Yuksel},
  \citenamefont {Kistler}, \citenamefont {Beacom},\ and\ \citenamefont
  {Hopkins}}]{Yuksel:2008cu}%
  \BibitemOpen
  \bibfield  {author} {\bibinfo {author} {\bibfnamefont {H.}~\bibnamefont
  {Yuksel}}, \bibinfo {author} {\bibfnamefont {M.~D.}\ \bibnamefont {Kistler}},
  \bibinfo {author} {\bibfnamefont {J.~F.}\ \bibnamefont {Beacom}}, \ and\
  \bibinfo {author} {\bibfnamefont {A.~M.}\ \bibnamefont {Hopkins}},\ }\href
  {\doibase 10.1086/591449} {\bibfield  {journal} {\bibinfo  {journal}
  {Astrophys.~J.}\ }\textbf {\bibinfo {volume} {683}},\ \bibinfo {pages} {L5}
  (\bibinfo {year} {2008})},\ \Eprint {http://arxiv.org/abs/0804.4008}
  {arXiv:0804.4008 [astro-ph]} \BibitemShut {NoStop}%
\bibitem [{\citenamefont {Anchordoqui}\ \emph {et~al.}(2014)\citenamefont
  {Anchordoqui} \emph {et~al.}}]{Anchordoqui:2013dnh}%
  \BibitemOpen
  \bibfield  {author} {\bibinfo {author} {\bibfnamefont {L.~A.}\ \bibnamefont
  {Anchordoqui}} \emph {et~al.},\ }\href {\doibase 10.1016/j.jheap.2014.01.001}
  {\bibfield  {journal} {\bibinfo  {journal} {JHEAp}\ }\textbf {\bibinfo
  {volume} {1-2}},\ \bibinfo {pages} {1} (\bibinfo {year} {2014})},\ \Eprint
  {http://arxiv.org/abs/1312.6587} {arXiv:1312.6587 [astro-ph.HE]} \BibitemShut
  {NoStop}%
\bibitem [{\citenamefont {Arg{\"u}elles}\ \emph {et~al.}(2015)\citenamefont
  {Arg{\"u}elles}, \citenamefont {Salvad{\'o}},\ and\ \citenamefont
  {Weaver}}]{Delgado:2014kpa}%
  \BibitemOpen
  \bibfield  {author} {\bibinfo {author} {\bibfnamefont {C.~A.}\ \bibnamefont
  {Arg{\"u}elles}}, \bibinfo {author} {\bibfnamefont {J.}~\bibnamefont
  {Salvad{\'o}}}, \ and\ \bibinfo {author} {\bibfnamefont {C.~N.}\ \bibnamefont
  {Weaver}},\ }\href {\doibase 10.1016/j.cpc.2015.06.022} {\bibfield  {journal}
  {\bibinfo  {journal} {Comput.\ Phys.\ Commun.}\ }\textbf {\bibinfo {volume}
  {196}},\ \bibinfo {pages} {569} (\bibinfo {year} {2015})},\ \Eprint
  {http://arxiv.org/abs/1412.3832} {arXiv:1412.3832 [hep-ph]} \BibitemShut
  {NoStop}%
\bibitem [{\citenamefont {Arg{\"u}elles}\ \emph {et~al.}(2018)\citenamefont
  {Arg{\"u}elles}, \citenamefont {Salvad{\'o}},\ and\ \citenamefont
  {Weaver}}]{SQuIDS}%
  \BibitemOpen
  \bibfield  {author} {\bibinfo {author} {\bibfnamefont {C.~A.}\ \bibnamefont
  {Arg{\"u}elles}}, \bibinfo {author} {\bibfnamefont {J.}~\bibnamefont
  {Salvad{\'o}}}, \ and\ \bibinfo {author} {\bibfnamefont {C.~N.}\ \bibnamefont
  {Weaver}},\ }\href@noop {} {}\bibinfo {howpublished}
  {\url{https://github.com/jsalvado/SQuIDS/}} (\bibinfo {year} {2018}),\
  \bibinfo {note} {{{SQuIDS}}}\BibitemShut {NoStop}%
\bibitem [{\citenamefont {Arg{\"u}elles}\ \emph {et~al.}(2019)\citenamefont
  {Arg{\"u}elles}, \citenamefont {Salvad{\'o}},\ and\ \citenamefont
  {Weaver}}]{NuSQuIDS}%
  \BibitemOpen
  \bibfield  {author} {\bibinfo {author} {\bibfnamefont {C.~A.}\ \bibnamefont
  {Arg{\"u}elles}}, \bibinfo {author} {\bibfnamefont {J.}~\bibnamefont
  {Salvad{\'o}}}, \ and\ \bibinfo {author} {\bibfnamefont {C.~N.}\ \bibnamefont
  {Weaver}},\ }\href@noop {} {}\bibinfo {howpublished}
  {\url{https://github.com/arguelles/nuSQuIDS}} (\bibinfo {year} {2019}),\
  \bibinfo {note} {{{NuSQuIDS}}}\BibitemShut {NoStop}%
\bibitem [{\citenamefont {Dziewonski}\ and\ \citenamefont
  {Anderson}(1981)}]{Dziewonski:1981xy}%
  \BibitemOpen
  \bibfield  {author} {\bibinfo {author} {\bibfnamefont {A.~M.}\ \bibnamefont
  {Dziewonski}}\ and\ \bibinfo {author} {\bibfnamefont {D.~L.}\ \bibnamefont
  {Anderson}},\ }\href {\doibase 10.1016/0031-9201(81)90046-7} {\bibfield
  {journal} {\bibinfo  {journal} {Phys.~Earth Planet.~Interiors}\ }\textbf
  {\bibinfo {volume} {25}},\ \bibinfo {pages} {297} (\bibinfo {year}
  {1981})}\BibitemShut {NoStop}%
\bibitem [{\citenamefont {Lu}(2019)}]{Lu:ICRC2019}%
  \BibitemOpen
  \bibfield  {author} {\bibinfo {author} {\bibfnamefont {L.}~\bibnamefont {Lu}}
  (\bibinfo {collaboration} {IceCube}),\ }\href@noop {} {\bibfield  {journal}
  {\bibinfo  {journal} {PoS}\ }\textbf {\bibinfo {volume} {ICRC2019}},\
  \bibinfo {pages} {945} (\bibinfo {year} {2019})}\BibitemShut {NoStop}%
\bibitem [{\citenamefont {Palomares-Ruiz}\ \emph {et~al.}(2015)\citenamefont
  {Palomares-Ruiz}, \citenamefont {Vincent},\ and\ \citenamefont
  {Mena}}]{Palomares-Ruiz:2015mka}%
  \BibitemOpen
  \bibfield  {author} {\bibinfo {author} {\bibfnamefont {S.}~\bibnamefont
  {Palomares-Ruiz}}, \bibinfo {author} {\bibfnamefont {A.~C.}\ \bibnamefont
  {Vincent}}, \ and\ \bibinfo {author} {\bibfnamefont {O.}~\bibnamefont
  {Mena}},\ }\href {\doibase 10.1103/PhysRevD.91.103008} {\bibfield  {journal}
  {\bibinfo  {journal} {Phys.\ Rev.\ D}\ }\textbf {\bibinfo {volume} {91}},\
  \bibinfo {pages} {103008} (\bibinfo {year} {2015})},\ \Eprint
  {http://arxiv.org/abs/1502.02649} {arXiv:1502.02649 [astro-ph.HE]}
  \BibitemShut {NoStop}%
\bibitem [{\citenamefont {Aartsen}\ \emph
  {et~al.}(2014{\natexlab{b}})\citenamefont {Aartsen} \emph
  {et~al.}}]{Aartsen:2013vja}%
  \BibitemOpen
  \bibfield  {author} {\bibinfo {author} {\bibfnamefont {M.~G.}\ \bibnamefont
  {Aartsen}} \emph {et~al.} (\bibinfo {collaboration} {IceCube}),\ }\href
  {\doibase 10.1088/1748-0221/9/03/P03009} {\bibfield  {journal} {\bibinfo
  {journal} {JINST}\ }\textbf {\bibinfo {volume} {9}},\ \bibinfo {pages}
  {P03009} (\bibinfo {year} {2014}{\natexlab{b}})},\ \Eprint
  {http://arxiv.org/abs/1311.4767} {arXiv:1311.4767 [physics.ins-det]}
  \BibitemShut {NoStop}%
\bibitem [{\citenamefont {Beacom}\ and\ \citenamefont
  {Candia}(2004)}]{Beacom:2004jb}%
  \BibitemOpen
  \bibfield  {author} {\bibinfo {author} {\bibfnamefont {J.~F.}\ \bibnamefont
  {Beacom}}\ and\ \bibinfo {author} {\bibfnamefont {J.}~\bibnamefont
  {Candia}},\ }\href {\doibase 10.1088/1475-7516/2004/11/009} {\bibfield
  {journal} {\bibinfo  {journal} {JCAP}\ }\textbf {\bibinfo {volume} {0411}},\
  \bibinfo {pages} {009} (\bibinfo {year} {2004})},\ \Eprint
  {http://arxiv.org/abs/hep-ph/0409046} {arXiv:hep-ph/0409046 [hep-ph]}
  \BibitemShut {NoStop}%
\bibitem [{\citenamefont {Feroz}\ and\ \citenamefont
  {Hobson}(2008)}]{Feroz:2007kg}%
  \BibitemOpen
  \bibfield  {author} {\bibinfo {author} {\bibfnamefont {F.}~\bibnamefont
  {Feroz}}\ and\ \bibinfo {author} {\bibfnamefont {M.~P.}\ \bibnamefont
  {Hobson}},\ }\href {\doibase 10.1111/j.1365-2966.2007.12353.x} {\bibfield
  {journal} {\bibinfo  {journal} {Mon.\ Not.\ Roy.\ Astron.\ Soc.}\ }\textbf
  {\bibinfo {volume} {384}},\ \bibinfo {pages} {449} (\bibinfo {year}
  {2008})},\ \Eprint {http://arxiv.org/abs/0704.3704} {arXiv:0704.3704
  [astro-ph]} \BibitemShut {NoStop}%
\bibitem [{\citenamefont {Feroz}\ \emph {et~al.}(2009)\citenamefont {Feroz},
  \citenamefont {Hobson},\ and\ \citenamefont {Bridges}}]{Feroz:2008xx}%
  \BibitemOpen
  \bibfield  {author} {\bibinfo {author} {\bibfnamefont {F.}~\bibnamefont
  {Feroz}}, \bibinfo {author} {\bibfnamefont {M.~P.}\ \bibnamefont {Hobson}}, \
  and\ \bibinfo {author} {\bibfnamefont {M.}~\bibnamefont {Bridges}},\ }\href
  {\doibase 10.1111/j.1365-2966.2009.14548.x} {\bibfield  {journal} {\bibinfo
  {journal} {Mon.\ Not.\ Roy.\ Astron.\ Soc.}\ }\textbf {\bibinfo {volume}
  {398}},\ \bibinfo {pages} {1601} (\bibinfo {year} {2009})},\ \Eprint
  {http://arxiv.org/abs/0809.3437} {arXiv:0809.3437 [astro-ph]} \BibitemShut
  {NoStop}%
\bibitem [{\citenamefont {Feroz}\ \emph {et~al.}(2013)\citenamefont {Feroz},
  \citenamefont {Hobson}, \citenamefont {Cameron},\ and\ \citenamefont
  {Pettitt}}]{Feroz:2013hea}%
  \BibitemOpen
  \bibfield  {author} {\bibinfo {author} {\bibfnamefont {F.}~\bibnamefont
  {Feroz}}, \bibinfo {author} {\bibfnamefont {M.~P.}\ \bibnamefont {Hobson}},
  \bibinfo {author} {\bibfnamefont {E.}~\bibnamefont {Cameron}}, \ and\
  \bibinfo {author} {\bibfnamefont {A.~N.}\ \bibnamefont {Pettitt}},\
  }\href@noop {} {\  (\bibinfo {year} {2013})},\ \Eprint
  {http://arxiv.org/abs/1306.2144} {arXiv:1306.2144 [astro-ph.IM]} \BibitemShut
  {NoStop}%
\bibitem [{\citenamefont {Buchner}\ \emph {et~al.}(2014)\citenamefont {Buchner}
  \emph {et~al.}}]{Buchner:2014nha}%
  \BibitemOpen
  \bibfield  {author} {\bibinfo {author} {\bibfnamefont {J.}~\bibnamefont
  {Buchner}} \emph {et~al.},\ }\href {\doibase 10.1051/0004-6361/201322971}
  {\bibfield  {journal} {\bibinfo  {journal} {Astron.\ Astrophys.}\ }\textbf
  {\bibinfo {volume} {564}},\ \bibinfo {pages} {A125} (\bibinfo {year}
  {2014})},\ \Eprint {http://arxiv.org/abs/1402.0004} {arXiv:1402.0004
  [astro-ph.HE]} \BibitemShut {NoStop}%
\bibitem [{\citenamefont {Haack}\ and\ \citenamefont
  {Wiebusch}(2018)}]{Haack:2017dxi}%
  \BibitemOpen
  \bibfield  {author} {\bibinfo {author} {\bibfnamefont {C.}~\bibnamefont
  {Haack}}\ and\ \bibinfo {author} {\bibfnamefont {C.}~\bibnamefont {Wiebusch}}
  (\bibinfo {collaboration} {IceCube}),\ }\href {\doibase 10.22323/1.301.1005}
  {\bibfield  {journal} {\bibinfo  {journal} {PoS}\ }\textbf {\bibinfo {volume}
  {ICRC2017}},\ \bibinfo {pages} {1005} (\bibinfo {year} {2018})}\BibitemShut
  {NoStop}%
\bibitem [{\citenamefont {Jeffreys}(1939)}]{Jeffreys:1939xee}%
  \BibitemOpen
  \bibfield  {author} {\bibinfo {author} {\bibfnamefont {H.}~\bibnamefont
  {Jeffreys}},\ }\href
  {https://global.oup.com/academic/product/the-theory-of-probability-9780198503682?cc=au&lang=en&#}
  {\emph {\bibinfo {title} {{The Theory of Probability}}}},\ Oxford Classic
  Texts in the Physical Sciences\ (\bibinfo {year} {1939})\BibitemShut
  {NoStop}%
\bibitem [{\citenamefont {Huang}\ and\ \citenamefont
  {Zhou}(2019)}]{Huang:2018nxj}%
  \BibitemOpen
  \bibfield  {author} {\bibinfo {author} {\bibfnamefont {G.-Y.}\ \bibnamefont
  {Huang}}\ and\ \bibinfo {author} {\bibfnamefont {S.}~\bibnamefont {Zhou}},\
  }\href {\doibase 10.1088/1475-7516/2019/02/024} {\bibfield  {journal}
  {\bibinfo  {journal} {JCAP}\ }\textbf {\bibinfo {volume} {02}},\ \bibinfo
  {pages} {024} (\bibinfo {year} {2019})},\ \Eprint
  {http://arxiv.org/abs/1810.03877} {arXiv:1810.03877 [hep-ph]} \BibitemShut
  {NoStop}%
\bibitem [{\citenamefont {Ando}(2003)}]{Ando:2003ie}%
  \BibitemOpen
  \bibfield  {author} {\bibinfo {author} {\bibfnamefont {S.}~\bibnamefont
  {Ando}},\ }\href {\doibase 10.1016/j.physletb.2003.07.009} {\bibfield
  {journal} {\bibinfo  {journal} {Phys.\ Lett.\ B}\ }\textbf {\bibinfo {volume}
  {570}},\ \bibinfo {pages} {11} (\bibinfo {year} {2003})},\ \Eprint
  {http://arxiv.org/abs/hep-ph/0307169} {arXiv:hep-ph/0307169 [hep-ph]}
  \BibitemShut {NoStop}%
\bibitem [{\citenamefont {Fogli}\ \emph {et~al.}(2004)\citenamefont {Fogli},
  \citenamefont {Lisi}, \citenamefont {Mirizzi},\ and\ \citenamefont
  {Montanino}}]{Fogli:2004gy}%
  \BibitemOpen
  \bibfield  {author} {\bibinfo {author} {\bibfnamefont {G.~L.}\ \bibnamefont
  {Fogli}}, \bibinfo {author} {\bibfnamefont {E.}~\bibnamefont {Lisi}},
  \bibinfo {author} {\bibfnamefont {A.}~\bibnamefont {Mirizzi}}, \ and\
  \bibinfo {author} {\bibfnamefont {D.}~\bibnamefont {Montanino}},\ }\href
  {\doibase 10.1103/PhysRevD.70.013001} {\bibfield  {journal} {\bibinfo
  {journal} {Phys.\ Rev.\ D}\ }\textbf {\bibinfo {volume} {70}},\ \bibinfo
  {pages} {013001} (\bibinfo {year} {2004})},\ \Eprint
  {http://arxiv.org/abs/hep-ph/0401227} {arXiv:hep-ph/0401227 [hep-ph]}
  \BibitemShut {NoStop}%
\bibitem [{\citenamefont {Beacom}\ \emph
  {et~al.}(2004{\natexlab{b}})\citenamefont {Beacom}, \citenamefont {Bell},\
  and\ \citenamefont {Dodelson}}]{Beacom:2004yd}%
  \BibitemOpen
  \bibfield  {author} {\bibinfo {author} {\bibfnamefont {J.~F.}\ \bibnamefont
  {Beacom}}, \bibinfo {author} {\bibfnamefont {N.~F.}\ \bibnamefont {Bell}}, \
  and\ \bibinfo {author} {\bibfnamefont {S.}~\bibnamefont {Dodelson}},\ }\href
  {\doibase 10.1103/PhysRevLett.93.121302} {\bibfield  {journal} {\bibinfo
  {journal} {Phys.\ Rev.\ Lett.}\ }\textbf {\bibinfo {volume} {93}},\ \bibinfo
  {pages} {121302} (\bibinfo {year} {2004}{\natexlab{b}})},\ \Eprint
  {http://arxiv.org/abs/astro-ph/0404585} {arXiv:astro-ph/0404585 [astro-ph]}
  \BibitemShut {NoStop}%
\bibitem [{\citenamefont {Serpico}(2007)}]{Serpico:2007pt}%
  \BibitemOpen
  \bibfield  {author} {\bibinfo {author} {\bibfnamefont {P.~D.}\ \bibnamefont
  {Serpico}},\ }\href {\doibase 10.1103/PhysRevLett.98.171301} {\bibfield
  {journal} {\bibinfo  {journal} {Phys.~Rev.~Lett.}\ }\textbf {\bibinfo
  {volume} {98}},\ \bibinfo {pages} {171301} (\bibinfo {year} {2007})},\
  \Eprint {http://arxiv.org/abs/astro-ph/0701699} {arXiv:astro-ph/0701699
  [astro-ph]} \BibitemShut {NoStop}%
\bibitem [{\citenamefont {Kachelriess}\ \emph {et~al.}(2000)\citenamefont
  {Kachelriess}, \citenamefont {Tomas},\ and\ \citenamefont
  {Valle}}]{Kachelriess:2000qc}%
  \BibitemOpen
  \bibfield  {author} {\bibinfo {author} {\bibfnamefont {M.}~\bibnamefont
  {Kachelriess}}, \bibinfo {author} {\bibfnamefont {R.}~\bibnamefont {Tomas}},
  \ and\ \bibinfo {author} {\bibfnamefont {J.~W.~F.}\ \bibnamefont {Valle}},\
  }\href {\doibase 10.1103/PhysRevD.62.023004} {\bibfield  {journal} {\bibinfo
  {journal} {Phys.~Rev.~D}\ }\textbf {\bibinfo {volume} {62}},\ \bibinfo
  {pages} {023004} (\bibinfo {year} {2000})},\ \Eprint
  {http://arxiv.org/abs/hep-ph/0001039} {arXiv:hep-ph/0001039 [hep-ph]}
  \BibitemShut {NoStop}%
\bibitem [{\citenamefont {Farzan}(2003)}]{Farzan:2002wx}%
  \BibitemOpen
  \bibfield  {author} {\bibinfo {author} {\bibfnamefont {Y.}~\bibnamefont
  {Farzan}},\ }\href {\doibase 10.1103/PhysRevD.67.073015} {\bibfield
  {journal} {\bibinfo  {journal} {Phys.~Rev.~D}\ }\textbf {\bibinfo {volume}
  {67}},\ \bibinfo {pages} {073015} (\bibinfo {year} {2003})},\ \Eprint
  {http://arxiv.org/abs/hep-ph/0211375} {arXiv:hep-ph/0211375 [hep-ph]}
  \BibitemShut {NoStop}%
\bibitem [{\citenamefont {Aartsen}\ \emph {et~al.}(2019)\citenamefont {Aartsen}
  \emph {et~al.}}]{Aartsen:2019swn}%
  \BibitemOpen
  \bibfield  {author} {\bibinfo {author} {\bibfnamefont {M.~G.}\ \bibnamefont
  {Aartsen}} \emph {et~al.} (\bibinfo {collaboration} {IceCube}),\ }\href@noop
  {} {\  (\bibinfo {year} {2019})},\ \Eprint {http://arxiv.org/abs/1911.02561}
  {arXiv:1911.02561 [astro-ph.HE]} \BibitemShut {NoStop}%
\bibitem [{\citenamefont {Iocco}\ \emph {et~al.}(2008)\citenamefont {Iocco},
  \citenamefont {Murase}, \citenamefont {Nagataki},\ and\ \citenamefont
  {Serpico}}]{Iocco:2007td}%
  \BibitemOpen
  \bibfield  {author} {\bibinfo {author} {\bibfnamefont {F.}~\bibnamefont
  {Iocco}}, \bibinfo {author} {\bibfnamefont {K.}~\bibnamefont {Murase}},
  \bibinfo {author} {\bibfnamefont {S.}~\bibnamefont {Nagataki}}, \ and\
  \bibinfo {author} {\bibfnamefont {P.~D.}\ \bibnamefont {Serpico}},\ }\href
  {\doibase 10.1086/526450} {\bibfield  {journal} {\bibinfo  {journal}
  {Astrophys.\ J.}\ }\textbf {\bibinfo {volume} {675}},\ \bibinfo {pages} {937}
  (\bibinfo {year} {2008})},\ \Eprint {http://arxiv.org/abs/0707.0515}
  {arXiv:0707.0515 [astro-ph]} \BibitemShut {NoStop}%
\bibitem [{\citenamefont {Giunti}\ and\ \citenamefont
  {Kim}(2007)}]{Giunti:2007ry}%
  \BibitemOpen
  \bibfield  {author} {\bibinfo {author} {\bibfnamefont {C.}~\bibnamefont
  {Giunti}}\ and\ \bibinfo {author} {\bibfnamefont {C.~W.}\ \bibnamefont
  {Kim}},\ }\href@noop {} {\emph {\bibinfo {title} {{Fundamentals of Neutrino
  Physics and Astrophysics}}}}\ (\bibinfo {year} {2007})\BibitemShut {NoStop}%
\bibitem [{\citenamefont {Dulat}\ \emph {et~al.}(2016)\citenamefont {Dulat}
  \emph {et~al.}}]{Dulat:2015mca}%
  \BibitemOpen
  \bibfield  {author} {\bibinfo {author} {\bibfnamefont {S.}~\bibnamefont
  {Dulat}} \emph {et~al.},\ }\href {\doibase 10.1103/PhysRevD.93.033006}
  {\bibfield  {journal} {\bibinfo  {journal} {Phys.\ Rev.\ D}\ }\textbf
  {\bibinfo {volume} {93}},\ \bibinfo {pages} {033006} (\bibinfo {year}
  {2016})},\ \Eprint {http://arxiv.org/abs/1506.07443} {arXiv:1506.07443
  [hep-ph]} \BibitemShut {NoStop}%
\bibitem [{\citenamefont {Mikaelian}\ and\ \citenamefont
  {Zheleznykh}(1980)}]{Mikaelian:1980vd}%
  \BibitemOpen
  \bibfield  {author} {\bibinfo {author} {\bibfnamefont {K.~O.}\ \bibnamefont
  {Mikaelian}}\ and\ \bibinfo {author} {\bibfnamefont {I.~M.}\ \bibnamefont
  {Zheleznykh}},\ }\href {\doibase 10.1103/PhysRevD.22.2122} {\bibfield
  {journal} {\bibinfo  {journal} {Phys.\ Rev.\ D}\ }\textbf {\bibinfo {volume}
  {22}},\ \bibinfo {pages} {2122} (\bibinfo {year} {1980})}\BibitemShut
  {NoStop}%
\bibitem [{\citenamefont {Gandhi}\ \emph {et~al.}(1996)\citenamefont {Gandhi},
  \citenamefont {Quigg}, \citenamefont {Reno},\ and\ \citenamefont
  {Sarcevic}}]{Gandhi:1995tf}%
  \BibitemOpen
  \bibfield  {author} {\bibinfo {author} {\bibfnamefont {R.}~\bibnamefont
  {Gandhi}}, \bibinfo {author} {\bibfnamefont {C.}~\bibnamefont {Quigg}},
  \bibinfo {author} {\bibfnamefont {M.~H.}\ \bibnamefont {Reno}}, \ and\
  \bibinfo {author} {\bibfnamefont {I.}~\bibnamefont {Sarcevic}},\ }\href
  {\doibase 10.1016/0927-6505(96)00008-4} {\bibfield  {journal} {\bibinfo
  {journal} {Astropart.\ Phys.}\ }\textbf {\bibinfo {volume} {5}},\ \bibinfo
  {pages} {81} (\bibinfo {year} {1996})},\ \Eprint
  {http://arxiv.org/abs/hep-ph/9512364} {arXiv:hep-ph/9512364} \BibitemShut
  {NoStop}%
\end{thebibliography}

%


\newpage
\clearpage

\appendix


\onecolumngrid

\begin{center}
 \large
 Supplemental Material for\\
 \smallskip
 {\it New limits on neutrino decay\\from the Glashow resonance of high-energy cosmic neutrinos}
\end{center}


\section{Flavor ratios with neutrino decay}

Because of the cosmological expansion, the energy of neutrinos emitted from redshift $z$ is a factor of $(1+z)$ smaller by the time they reach Earth, which affects their lifetime in the lab frame.  We account for this by following \Ref\ \cite{Baerwald:2012kc} to compute the fraction of unstable mass eigenstates $\nu_j$ that remains upon reaching Earth as\
$D = \left[ \mathcal{Z}\left(z\right) \right]^{- \frac{m_j}{\tau_j} \cdot \frac{L_{\rm H}}{E_\nu}}$, where $L_{\rm H} \approx 3.89$ Gpc is the Hubble length, $\mathcal{Z}\left(z\right) \simeq a + b e^{-cz}$, $a \approx1.71$, $b = 1-a$, and $c \approx 1.27$ for a $\Lambda$CDM cosmology with $\Omega_m = 0.27$ and $\Omega_\Lambda = 0.73$ the adimensional energy densities of matter and vacuum, respectively.  For stable eigenstates, $D = 1$; for unstable ones, $D < 1$. If $D \ll 1$ for all unstable eigenstates, decay is complete.

The decay of $\nu_j$ into $\nu_3$ changes the flavor ratio $f_{\alpha,\oplus}$ by a net factor $\propto (\lvert U_{\alpha j} \rvert^2 - \lvert U_{\alpha 3} \rvert^2) D$.  
Via $D$, the flavor ratios at Earth acquire a dependence on $E_\nu$, $z$, $\tau_1/m_1$, and $\tau_2/m_2$.  Reference\ \cite{Bustamante:2016ciw} derived a general expression to compute them.  In the inverted neutrino mass ordering, this is
\begin{equation*}
 f_{\alpha,\oplus} \left( E_\nu, z; f_{\beta, {\rm S}}, \boldsymbol\theta, \frac{\tau_1}{m_1}, \frac{\tau_2}{m_2} \right)
 =
 \lvert U_{\alpha 3} (\boldsymbol\theta) \rvert^2 
 + \sum_{j \neq 3}
 f_{j,\text{S}} ( f_{\beta, {\rm S}}, \boldsymbol\theta )
 \left( 
 \lvert U_{\alpha j} (\boldsymbol\theta) \rvert^2 - \lvert U_{\alpha 3} (\boldsymbol\theta) \rvert^2
 \right)
 D\left(E_\nu,z,\frac{\tau_j}{m_j}\right) \;,
\end{equation*}
where $f_{\beta, {\rm S}}$ are the flavor ratios at the source and $\boldsymbol\theta \equiv (\theta_{12}, \theta_{23}, \theta_{13}, \delta_{\rm CP})$.  
The mass-eigenstate ratios are computed from the flavor ratios as $f_{j,\text{S}} = \sum_\alpha f_{\alpha,\text{S}} \left\vert U_{\alpha j} (\boldsymbol\theta) \right\vert^2$.  

Figure\ \ref{fig:flavor_ratios}, left panel, illustrates the evolution of $f_{e, \oplus}$ with neutrino energy, for fixed redshift, lifetime, and mixing parameters, assuming the nominal expectation of flavor ratios at the source.  At high energies, where the lifetimes in the lab frame are longer, $f_{e, \oplus} \approx 1/3$.  At low energies, where the lifetimes in the lab frame are shorter, decay is complete.  There, when $\nu_1$ and $\nu_2$ are unstable, $f_{e, \oplus} = \left\vert U_{e3} \right\vert^2 \approx 0.02$.  At low energies, it is evident that the decay of $\nu_1$ affects the electron-flavor content of the neutrino flux the most, which is why our analysis is mainly sensitive to $\nu_1$.

Figure\ \ref{fig:flavor_ratios}, right panel, illustrates the evolution with neutrino lifetime of $f_{e, \oplus}$ computed at the GR energy of 6.3~PeV.  It shows that the GR, and, thus, our analysis, is sensitive to neutrino lifetimes smaller than about $10^3$~s~eV$^{-1}$.


\section{Diffuse flux of high-energy cosmic neutrinos}

The diffuse energy flux of $\nu_\alpha$ at Earth is (see, \eg, \Refs\ \cite{Iocco:2007td, Murase:2015xka, Bustamante:2016ciw})
\begin{equation*} 
 E_\nu^2 \Phi_{\nu_\alpha}^\oplus
 \left( E_\nu; \phi_0, \gamma, f_{e, {\rm S}}, f_{\mu, {\rm S}}, f_{\bar{\nu}}, \boldsymbol\theta, \frac{\tau_1}{m_1}, \frac{\tau_2}{m_2} \right) 
 =
 \frac{L_{\rm H}}{4\pi} 
 \int_0^{z_{\max}} dz
 \frac{\rho_{\rm src}(z)}{h(z)(1+z)^2}
 J_{\nu_\alpha} \left(E_\nu, z; \phi_0, \gamma, f_{e, {\rm S}}, f_{\mu, {\rm S}}, f_{\bar{\nu}}, \boldsymbol\theta, \frac{\tau_1}{m_1}, \frac{\tau_2}{m_2}\right)
 \;,
\end{equation*}
where  $z_{\max} = 4$, and $h(z) \equiv [\Omega_m (1+z)^3 + \Omega_\Lambda]^{1/2}$ is the adimensional Hubble parameter.  The contribution of $\nu_{\alpha}$ from sources at redshift $z$ is 
\begin{equation*}
 J_{\nu_\alpha} \left(E_\nu, z; \phi_0, \gamma, f_{e, {\rm S}}, f_{\mu, {\rm S}}, f_{\bar{\nu}}, \boldsymbol\theta, \frac{\tau_1}{m_1}, \frac{\tau_2}{m_2} \right)  
 =
 \phi_0 [(1+z) E_\nu]^{2-\gamma} (1-f_{\bar{\nu}}) 
 f_{\alpha, \oplus} \left(E_\nu, z; f_{e, {\rm S}}, f_{\mu, {\rm S}}, \boldsymbol\theta, \frac{\tau_1}{m_1}, \frac{\tau_2}{m_2} \right) \;.
\end{equation*}
The flux $E_\nu^2 \Phi_{\bar{\nu}_\alpha}^\oplus$ of $\bar{\nu}_\alpha$ is the same as above, with $(1-f_{\bar{\nu}})$ replaced by $f_{\bar{\nu}}$.  The redshift evolution of the neutrino luminosity density $\rho_{\rm src} J_{\nu_\alpha}$ follows the star formation rate\ \cite{Yuksel:2008cu}.

\setcounter{figure}{0}
\renewcommand{\thefigure}{A\arabic{table}}
\begin{figure}[t!]
 \centering
 \includegraphics[width=0.495\columnwidth,height=0.495\columnwidth]{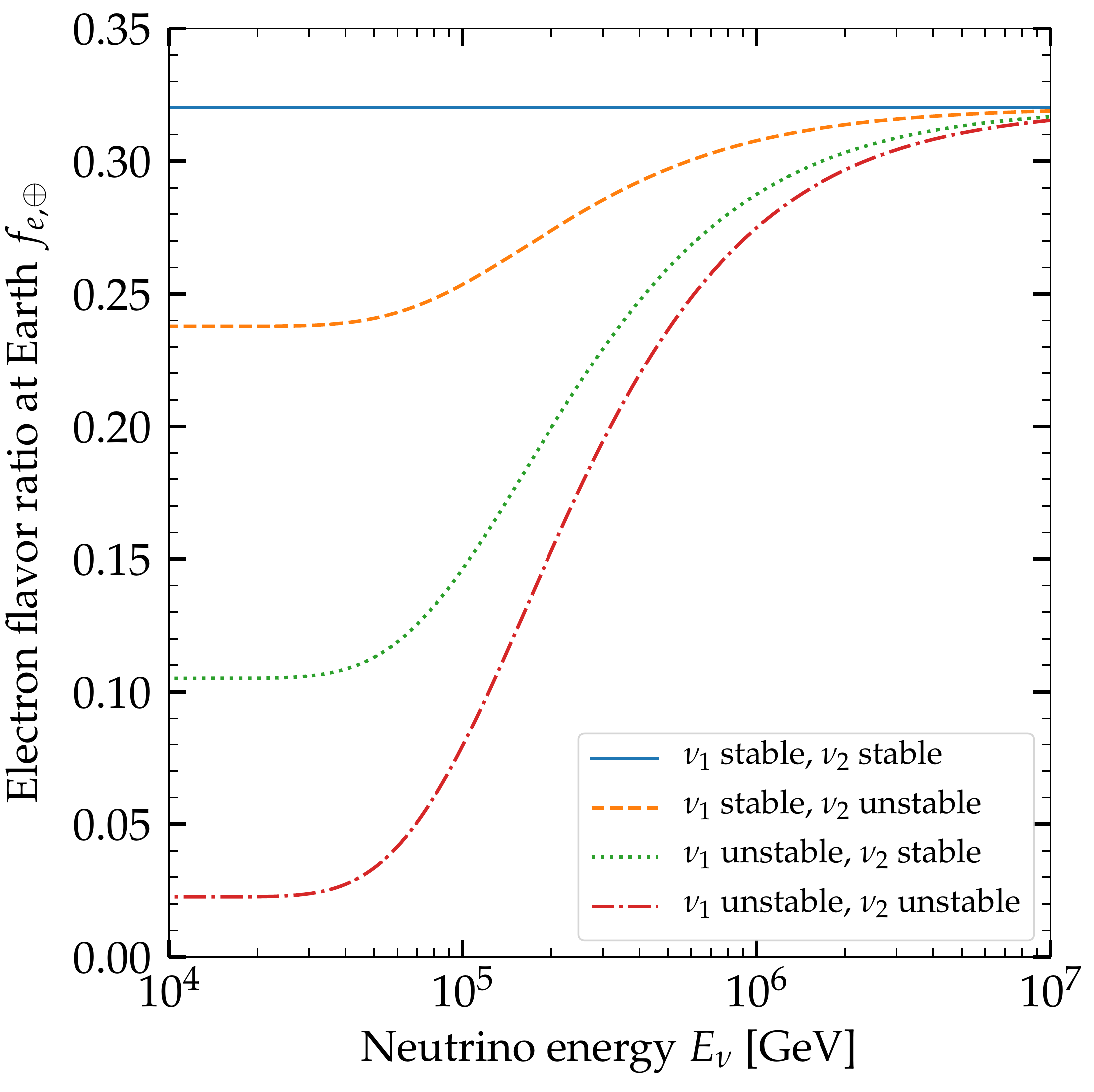}
 \includegraphics[width=0.495\columnwidth,height=0.495\columnwidth]{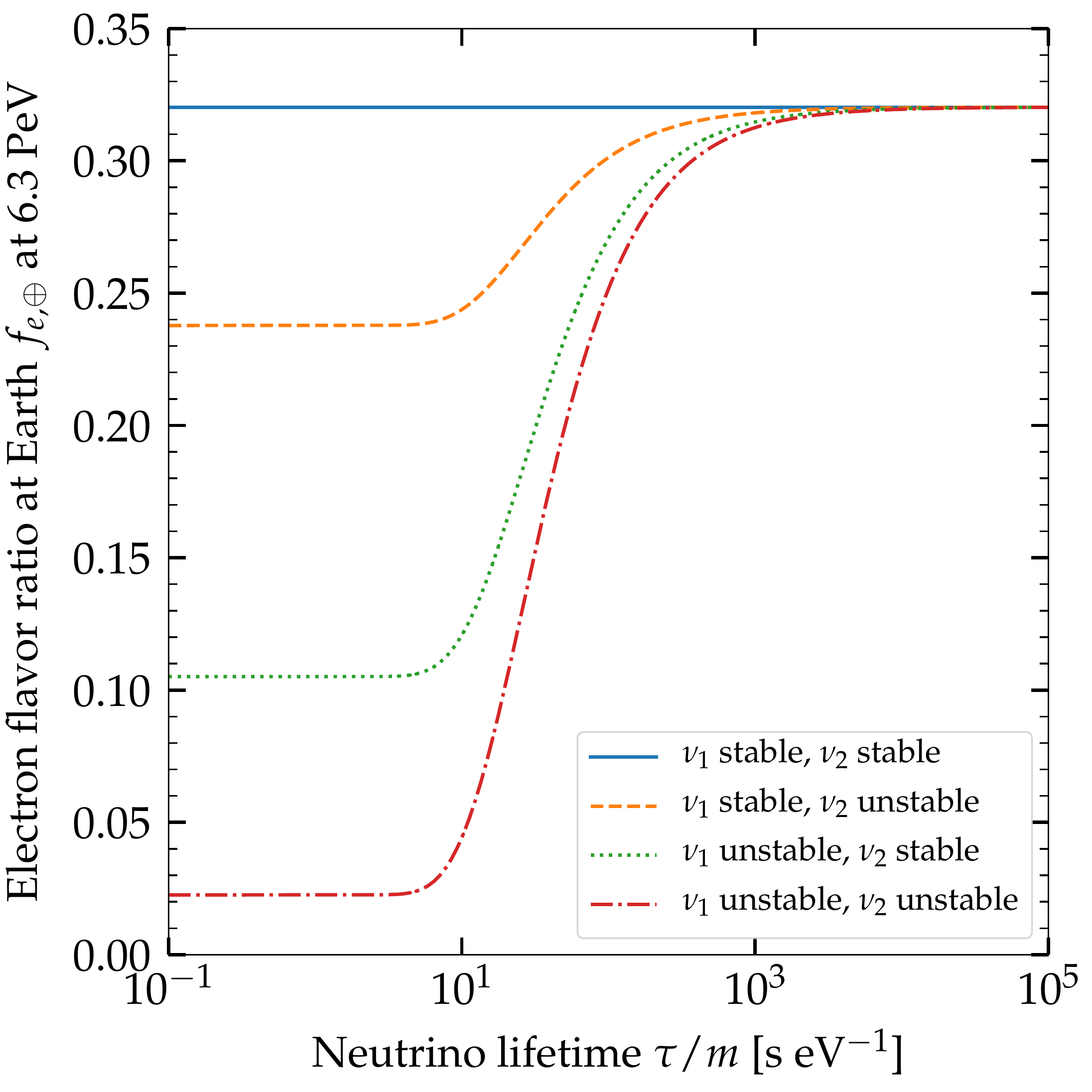}
 \caption{\label{fig:flavor_ratios}
 Electron flavor ratio at Earth, $f_{e, \oplus}$, allowing for the decay of $\nu_1$ and $\nu_2$ into $\nu_3$.  For this plot, for the purpose of illustration, we fix the source redshift to $z = 1$, the flavor ratios at the source to $\left( f_e : f_\mu : f_\tau \right)_{\rm S} = \left( \frac{1}{3} : \frac{2}{3} : 0 \right)$, and the mixing parameters to their best-fit values from {\tt NuFit}~4.1 in the inverted mass ordering\ \cite{deSalas:2018bym, NuFit_4.1}.  We show separately results for one unstable neutrino, two unstable neutrinos (with equal lifetime), or no unstable neutrino; in our analysis, we always allow both neutrinos to be unstable.  {\it Left:}  Variation of $f_{e, \oplus}$ with neutrino energy.  Unstable neutrinos have a fixed lifetime of $10^3$~s~eV$^{-1}$.  {\it Right:}  Variation of $f_{e, \oplus}$ evaluated at the GR energy of 6.3~PeV with lifetime.  
 }
\end{figure}

\setcounter{figure}{0}
\renewcommand{\thefigure}{B\arabic{table}}
\begin{figure}[t!]
 \centering
 \includegraphics[width=0.5\columnwidth,height=0.5\columnwidth]{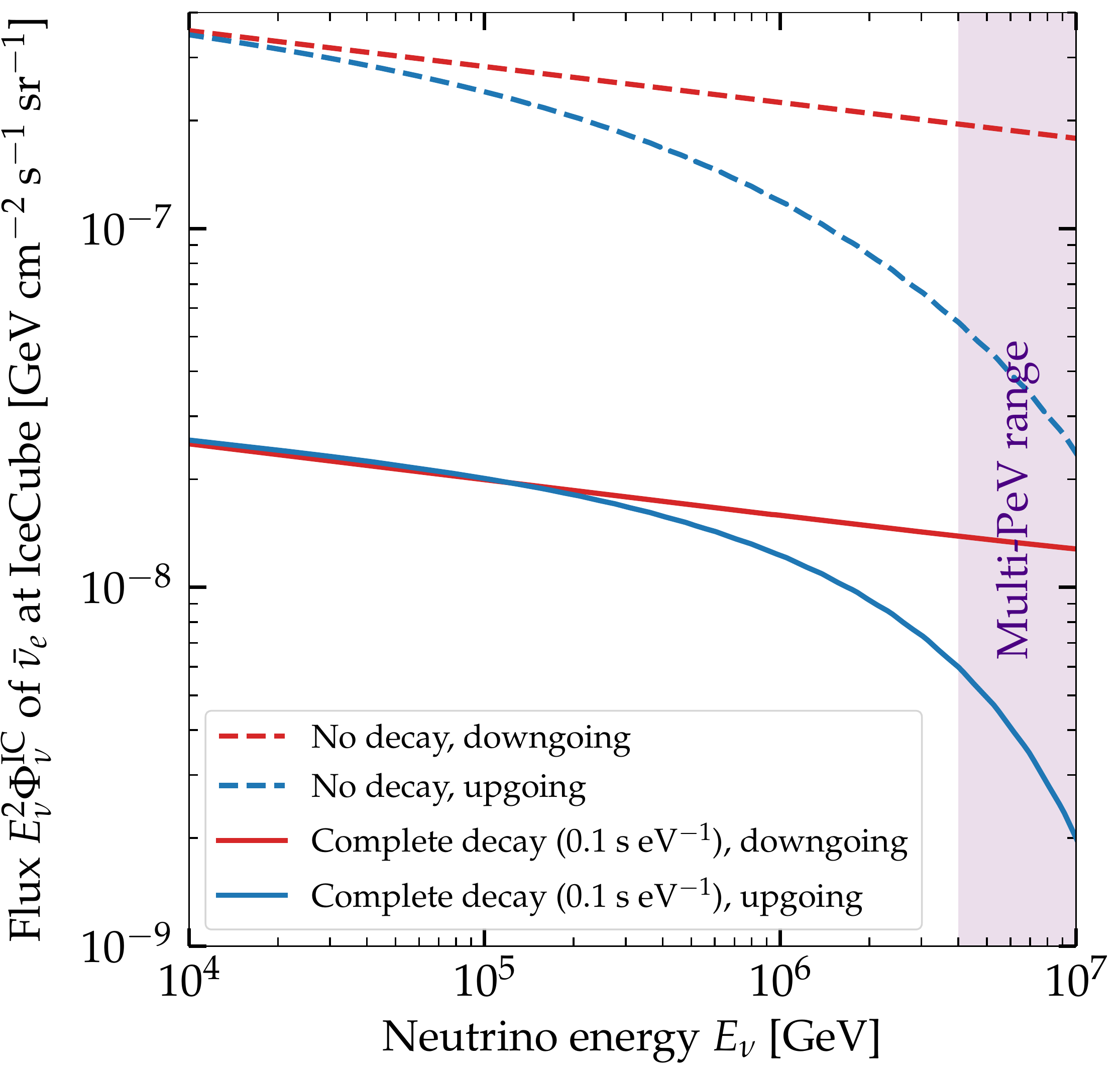}
 \caption{\label{fig:flux_decay}
 Diffuse flux of high-energy cosmic $\bar{\nu}_e$ that reaches IceCube without decay and with complete decay, for two  directions: downgoing ($\cos \theta_z = 1$) and upgoing ($\cos \theta_z = -0.55$).  For this plot, we choose the same illustrative parameters values as in \figu{shower_spectrum}.  In our analysis, we let these parameters vary; see the main text.}
\end{figure}

Figure\ \ref{fig:flux_decay} compares the flux of $\bar{\nu}_e$, without and with decay, that arrives at IceCube after propagating inside the Earth, for an illustrative choice of values of the free parameters.  
It shows that in-Earth propagation only affects the flux of upgoing neutrinos ($\cos \theta_z < 0$).


\section{Computation of shower rates}

The number of contained showers detected in IceCube with deposited energies $E_{\rm dep} = 4$--8~PeV in a time $T$, integrated over all arrival directions, is
\begin{eqnarray*}
 &&
 N_{\rm sh}
 \left( \phi_0, \gamma, f_{e, {\rm S}}, f_{\mu, {\rm S}}, f_{\bar{\nu}}, \boldsymbol\theta, \frac{\tau_1}{m_1}, \frac{\tau_2}{m_2} \right)
 \nonumber \\
 && =
 2\pi T 
 \int_{4~{\rm PeV}}^{8~{\rm PeV}} dE_{\rm dep}
 \int_{-1}^1 d\cos\theta_z
 \frac{dN_{\rm sh}}{dE_{\rm dep}} \left( E_{\rm dep}, \cos\theta_z;
 \phi_0, \gamma, f_{e, {\rm S}}, f_{\mu, {\rm S}}, f_{\bar{\nu}}, \boldsymbol\theta, \frac{\tau_1}{m_1}, \frac{\tau_2}{m_2} \right) \;.
\end{eqnarray*}
The shower spectrum $dN_{\rm sh}/dE_{\rm dep}$ is made up of contributions from all flavors, \ie,
\begin{equation*}
 \frac{dN_{\rm sh}}{dE_{\rm dep}}
 =
 \frac{dN_{{\rm sh}, e}}{dE_{\rm dep}} 
 +
 \frac{dN_{{\rm sh}, \mu}}{dE_{\rm dep}}
 +
 \frac{dN_{{\rm sh}, \tau}}{dE_{\rm dep}}
 \;,
\end{equation*}
where the contribution of each flavor, from $\nu N$ NC and CC interactions, and from the GR, is
\begin{eqnarray*}
 \frac{dN_{{\rm sh}, e}}{dE_{\rm dep}}
 &=&
 \frac{dN_{{\rm sh}, \nu_e}^{\rm NC}}{dE_{\rm dep}}
 +
 \frac{dN_{{\rm sh}, \nu_e}^{\rm CC}}{dE_{\rm dep}}
 +
 \frac{dN_{{\rm sh}, \bar{\nu}_e}^{\rm NC}}{dE_{\rm dep}}
 +
 \frac{dN_{{\rm sh}, \bar{\nu}_e}^{\rm CC}}{dE_{\rm dep}}
 +
 \frac{dN_{{\rm sh}, \bar{\nu}_e}^{\rm GR}}{dE_{\rm dep}}
 \;, \\
 \frac{dN_{{\rm sh}, \mu}}{dE_{\rm dep}}
 &=&
 \frac{dN_{{\rm sh}, \nu_\mu}^{\rm NC}}{dE_{\rm dep}}
 +
 \frac{dN_{{\rm sh}, \nu_\mu}^{\rm CC}}{dE_{\rm dep}}
 +
 \frac{dN_{{\rm sh}, \bar{\nu}_\mu}^{\rm NC}}{dE_{\rm dep}}
 +
 \frac{dN_{{\rm sh}, \bar{\nu}_\mu}^{\rm CC}}{dE_{\rm dep}}
 \;, \\
 \frac{dN_{{\rm sh}, \tau}}{dE_{\rm dep}}
 &=&
 \frac{dN_{{\rm sh}, \nu_\tau}^{\rm NC}}{dE_{\rm dep}}
 +
 \frac{dN_{{\rm sh}, \nu_\tau}^{\rm CC}}{dE_{\rm dep}}
 +
 \frac{dN_{{\rm sh}, \bar{\nu}_\tau}^{\rm NC}}{dE_{\rm dep}}
 +
 \frac{dN_{{\rm sh}, \bar{\nu}_\tau}^{\rm CC}}{dE_{\rm dep}}
 \; .
\end{eqnarray*}

To compute the shower spectra coming from the different interaction channels, we follow \Ref\ \cite{Palomares-Ruiz:2015mka}.  Below, we only outline the computation; for details, see \Ref\ \cite{Palomares-Ruiz:2015mka}.  As illustration, the shower spectrum due to NC interactions of the flux $\Phi_{\nu_\alpha}^{\rm IC} \equiv dN_\nu/dE_\nu$ of $\nu_\alpha$ that arrives at IceCube from the direction $\cos \theta_z$, after propagating inside the Earth, is
\begin{eqnarray*}
 &&
 \frac{dN_{\nu_\alpha}^{{\rm sh}, {\rm NC}}(E_{\rm dep}, \cos\theta_z)}{dE_{\rm dep}}
 \\
 &&
 =
 N_{\rm A}
 \int_0^\infty
 dE_\nu
 \Phi_{\nu_\alpha}^{\rm IC}(E_\nu, \cos \theta_z)
 \int_0^1
 dy
 M_{\rm eff}(E_{\rm true}(E_\nu))
 R(E_{\rm true}(E_\nu), E_{\rm dep}, \sigma(E_{\rm true}(E_\nu))
 \frac{d\sigma_{\nu_\alpha}^{\rm NC}(E_\nu, y)}{dy} \;.
\end{eqnarray*}
Here, $N_{\rm A} = 6.022 \times 10^{-23}$~g$^{-1}$ is Avogadro's number, $M_{\rm eff}$ is the effective IceCube mass\ \cite{Palomares-Ruiz:2015mka}, and $\sigma_{\nu_\alpha}^{\rm NC}$ is the $\nu N$ NC cross section.  The inelasticity $y$ is the fraction of the neutrino energy given to the final-state hadrons; the final-state lepton receives the remaining fraction $(1-y)$.  The energy resolution function $R$ describes the mismatch between the measured deposited energy, $E_{\rm dep}$ and the true deposited energy, $E_{\rm true}$, which varies with $E_\nu$.  It is a Gaussian with a spread $\sigma \approx 0.12 E_{\rm true}$\ \cite{Palomares-Ruiz:2015mka}.  The contribution of $\bar{\nu}_\alpha$ is the same as above, with $\nu_\alpha \to \bar{\nu}_\alpha$.  

For the $\nu N$ CC interactions of $\nu_e$ and $\nu_\tau$, the expressions are similar to the one above, with NC $\to$ CC.  However, for $\nu_\tau$, we compute separately the contribution of each decay channel of the final-state tauon.  The relation between $E_{\rm true}$ and $E_\nu$ depends on the flavor, interaction channel, and decay channel of final-state unstable particles\ \cite{Palomares-Ruiz:2015mka}.  Because in CC interactions of $\nu_e$ and $\nu_\tau$ all final-state particles shower, $E_{\rm dep}$ traces $E_\nu$ more closely than in NC interactions, where only the final-state hadrons shower.

For the shower rate from $\nu N$ interactions, we build the differential deep-inelastic-scattering cross sections on protons and neutrons\ \cite{Giunti:2007ry}, $d\sigma_{p, \nu_\alpha}^{\rm NC} / dy$, $d\sigma_{n, \nu_\alpha}^{\rm NC} / dy$, and their CC equivalents, using the recent CTEQ14 parton distribution functions\ \cite{Dulat:2015mca}, for $\nu_\alpha$ and $\bar{\nu}_\alpha$.  We weight these cross sections by the mass number $A = 18$, atomic number $Z = 10$, and neutron number $N = 8$ of water, \ie,
\begin{equation*}
 \frac{d\sigma_{\nu_\alpha}^{\rm NC}}{dy}
 =
 \frac{1}{A}
 \left(
 Z
 \frac{d\sigma_{p, \nu_\alpha}^{\rm NC}}{dy}
 +
 N
 \frac{d\sigma_{n, \nu_\alpha}^{\rm NC}}{dy}
 \right) \;,
\end{equation*}
and similarly for CC interactions.

For the shower rate from the Glashow resonance, we compute the $\bar{\nu}_e e$ differential cross section following \Refs\ \cite{Mikaelian:1980vd, Gandhi:1995tf}.  Following \Ref\ \cite{Palomares-Ruiz:2015mka}, we compute separately the contributions of each decay channel of the $W$ boson created in the resonance, into electrons, tauons, and hadrons.  For the tauonic decay, the computation accounts for the fraction of high-energy tauons that escape the detector volume before decaying, thus not contributing to the shower rate.


\section{Details of the statistical analysis}

Given the likelihood function $\mathcal{L}$, defined in the main text, and for a number $N_{\rm obs}$ of observed multi-PeV showers, we adopt a Bayesian approach and maximize the posterior probability distribution 
\begin{eqnarray*}
 \mathcal{P} \left( \phi_0, \gamma, f_{e, {\rm S}}, f_{\mu, {\rm S}}, f_{\bar{\nu}, \boldsymbol\theta}, \frac{\tau_1}{m_1}, \frac{\tau_2}{m_2} ; N_{\rm obs} \right)
 =
 &&
 ~
 \mathcal{P}(\phi_0)
 \mathcal{P}(\gamma)
 \mathcal{P}(f_{e, {\rm S}})
 \mathcal{P}(f_{\mu, {\rm S}})
 \mathcal{P}(\boldsymbol\theta)
 \mathcal{P}\left(\frac{\tau_1}{m_1}\right)
 \mathcal{P}\left(\frac{\tau_2}{m_2}\right)
 \\
 &&
 \times~
 \mathcal{L}\left( \phi_0, \gamma, f_{e, {\rm S}}, f_{\mu, {\rm S}}, f_{\bar{\nu}}, \boldsymbol\theta, \frac{\tau_1}{m_1}, \frac{\tau_2}{m_2}; N_{\rm obs}\right)
 \;.
\end{eqnarray*}
Here, $\mathcal{P}(p)$ is the prior probability distribution for the parameter $p$, and we abbreviate $\mathcal{P}(\boldsymbol\theta) \equiv \mathcal{P}(s_{12}) \mathcal{P}(s_{23}) \mathcal{P}(s_{13}) \mathcal{P}(\delta_{\rm CP})$.  To place constraints on any one parameter, we marginalize over all the others.   For the lifetime $\tau_1/m_1$, this is
\begin{equation*}
 \mathcal{P}_{\rm marg} \left(\frac{\tau_1}{m_1}; N_{\rm obs}\right)
 =
 \int d\phi_0
 \int d\gamma
 \int_0^1 df_{e, {\rm S}}
 \int_0^{1-f_{e, {\rm S}}} df_{\mu, {\rm S}}
 \int df_{\bar{\nu}}
 \int d\boldsymbol\theta
 \int d\frac{\tau_2}{m_2}
 ~\mathcal{P} \left( \phi_0, \gamma, f_{e, {\rm S}}, f_{\mu, {\rm S}}, f_{\bar{\nu}, \boldsymbol\theta}, \frac{\tau_1}{m_1}, \frac{\tau_2}{m_2} ; N_{\rm obs} \right) \;.
\end{equation*}
To find the one-dimensional allowed range of values of $\tau_1/m_1$ at the 68\%, 90\%, and $3\sigma$ credible intervals, we integrate this marginalized posterior, starting from the point where the posterior is maximum, until the above integral is a fraction of the total volume equal to the desired credibility level.  To find the allowed range of values of $\tau_2/m_2$, we integrate instead over $\tau_1/m_1$ in the expression above.  We use {\tt MultiNest}\ \cite{Feroz:2007kg, Feroz:2008xx, Feroz:2013hea, Buchner:2014nha}, an efficient implementation of the multimodal importance nested sampling algorithm for Bayesian analysis, to explore the large parameter space, and to find the maximum value of the posterior and the credible intervals.

Figure\ \ref{fig:pdf_mix_params} shows the priors of the mixing parameters, built from {\tt NuFit}~4.1 one-dimensional $\chi^2$ distributions\ \cite{deSalas:2018bym, NuFit_4.1}, assuming the inverted neutrino mass ordering, and using Super-Kamiokande atmospheric data.

Table \ref{tab:fit_parameters} shows, for each of the free parameters, their priors and their present-day (\ie, $N_{\rm obs } = 1$) allowed posterior ranges.  The allowed ranges of all of the  parameters agree comfortably with theory expectations. 

\setcounter{figure}{0}
\renewcommand{\thefigure}{D\arabic{figure}}
\begin{figure}[t!]
 \centering
 \includegraphics[width=0.5\columnwidth,height=0.5\columnwidth]{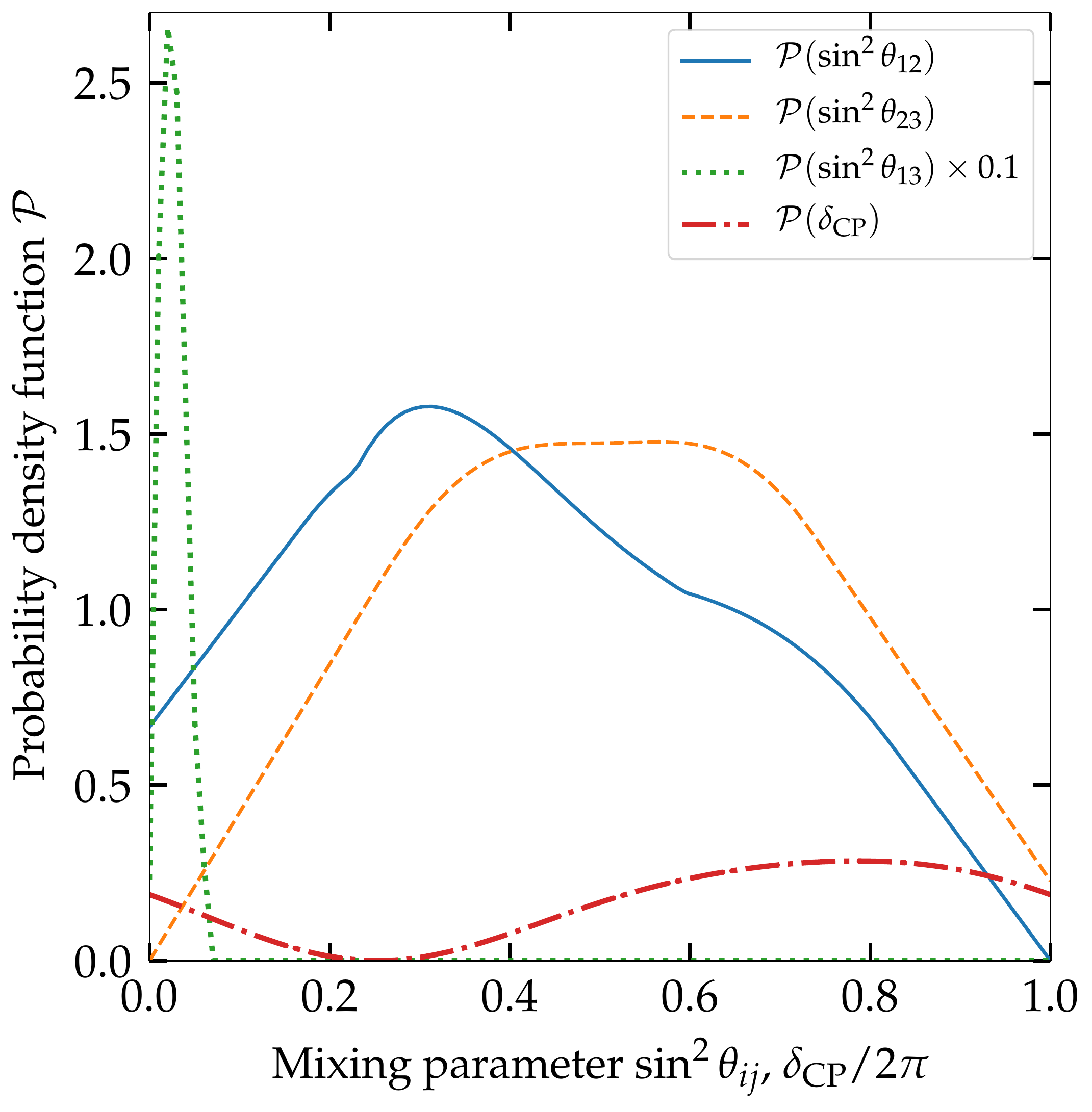}
 \caption{\label{fig:pdf_mix_params}Probability density functions of the mixing parameters, built from {\tt NuFit}~4.1 one-dimensional $\chi^2$ distributions\ \cite{deSalas:2018bym, NuFit_4.1}.}
\end{figure}

\setcounter{table}{0}
\renewcommand{\thetable}{D\arabic{table}}
\begin{table*}[t!]
 \begin{ruledtabular}
  \caption{\label{tab:fit_parameters}Parameters varied in the statistical analysis, their priors, and their posterior allowed ranges.  For the per-species normalization of the neutrino flux at 100~TeV ($\phi_0$) and spectral index ($\gamma$), the priors are built from recent 8-year IceCube $\nu_\mu$ diffuse flux\ \cite{Haack:2017dxi}.  For the flavor composition at the sources ($f_{e,{\rm S}}$, $f_{\mu,{\rm S}}$) and the fraction of $\bar{\nu}$ in the flux ($f_{\bar{\nu}}$), the priors cover their full allowed range of values.  For the mixing parameters ($\sin \theta_{12}$, $\sin \theta_{23}$, $\sin \theta_{13}$, $\delta_{\rm CP}$), the priors are built from their one-dimensional $\chi^2$ profiles from the {\tt NuFit}~4.1 global fit to oscillation data\ \cite{deSalas:2018bym, NuFit_4.1}; see \figu{pdf_mix_params}.  For the lifetimes of $\nu_1$ and $\nu_2$ ($\tau_1/m_1$, $\tau_2/m_2$), the priors are wide to avoid bias, and cover the range of values where high-energy cosmic neutrinos are sensitive.  The posterior allowed ranges of the parameters are extracted from the observation of $N_{\rm obs} = 1$ shower in the range 4--8~PeV in 4.6~years of IceCube.  For each parameter, the range shown is marginalized over all other parameters.}
  \centering
  \begin{tabular}{ccccc}
   \multirow{2}{*}{Parameter} & \multirow{2}{*}{Prior} & \multicolumn{3}{c}{Posterior allowed range} \\
            &        & Best fit $\pm 1\sigma$ & $90\% ~{\rm C.L.}$ & $3\sigma$ \\
   \hline
   $\phi_0~[10^{-8}~{\rm GeV}~{\rm cm}^{-2}~{\rm s}^{-1}~{\rm sr}^{-1}]$ & Normal on $9.00 \pm 2.05$  & $9.36 \pm 1.78$ & $[6.42,12.30]$ & $[3.39,14.97]$      \\
   $\gamma$                                                      & Normal on $2.19 \pm 0.10$          & $2.12 \pm 0.07$ & $[2.00,2.22]$ & $[1.87,2.37]$     \\
   $f_{e,{\rm S}}$                                               & Uniform in $[0,1]$                 & $0.56 \pm 0.31$ & $[0.08,0.94]$ & $[0.00,1.00]$   \\
   $f_{\mu,{\rm S}}$                                             & Uniform in $[0,1-f_{e,{\rm S}}]$   & $0.17 \pm 0.19$ & $[0.01,0.63]$ & $[0.00, 0.92]$  \\
   $f_{\bar{\nu}}$                                               & Uniform in $[0,1]$                 & $0.66 \pm 0.26$ & $[0.20,0.96]$ & $[0.02,1.00]$              \\
   $\sin \theta_{12}$                                            & From {\tt NuFit}~4.1 profile      & $0.62 \pm 0.20$ & $[0.29,0.89]$ & $[0.07,0.98]$  \\ 
   $\sin \theta_{23}$                                            & From {\tt NuFit}~4.1 profile      & $0.72 \pm 0.16$ & $[0.42,0.93]$ & $[0.18,1.00]$  \\
   $\sin \theta_{13}$                                            & From {\tt NuFit}~4.1 profile      & $0.16 \pm 0.04$ & $[0.08,0.22]$ & $[0.01,0.26]$  \\
   $\delta_{\rm CP}~[^\circ]$                                    & From {\tt NuFit}~4.1 profile      & $119.90 \pm 18.76$ & $[38.13,140.10]$ & $[7.68,143.46]$ \\
   
   $\log_{10}(\frac{\tau_1 / m_1}{{\rm s~eV}^{-1}})$             & Uniform in $[-5,15]$               & $8.03 \pm 5.22$  & $> -2.54$  & $> -4.94$ \\       
   
   $\log_{10}(\frac{\tau_2 / m_2}{{\rm s~eV}^{-1}})$             & Uniform in $[-5,15]$               & $6.81 \pm 5.77$  & $> -2.90$ & $> -5.00$ \\
   \end{tabular}
 \end{ruledtabular}
\end{table*}


\section{Projections for IceCube and IceCube-Gen2}

Figure\ \ref{fig:limits_future} shows present-day and projected lower limits on the lifetimes of $\nu_1$ and $\nu_2$, assuming they decay to a visible $\nu_3$, \ie, in the inverted neutrino mass ordering; see the main text for details.  The limits come from the observation of $N_{\rm obs}$ showers in the 4--8~PeV energy range in IceCube, dominated by showers made by the GR.

Our present-day limits come from having observed $N_{\rm obs} = 1$ shower in 4.6~years of IceCube, the first GR candidate\ \cite{Talk_Lu_UHECR_2018}; see \figu{limits} in the main text.  Projected IceCube limits come from observing $N_{\rm obs} = 2$ (9.2 years), 3 (13.8 years), and 4 event (18.4 years).  Projected limits with IceCube-Gen2 come from observing 4 events in IceCube plus 2 events in IceCube-Gen2 (2 years), and were obtained by fixing the mixing parameters to their current best-fit values from {\tt NuFit}~4.1, assuming inverted mass ordering using the Super-Kamiokande atmospheric data\ \cite{deSalas:2018bym, NuFit_4.1}.

Figure\ \ref{fig:limits_g_m} shows present-day and projected upper limits on the combined scalar ($g_{j3}$, with $j=1,2$) and pseudoscalar ($h_{j3}$) couplings of $\nu_1$ and $\nu_2$ to a new, light boson $\phi$ into which they decay, \ie, $(g_{j3}^2+h_{j3}^2)^{1/2}$, inferred from the limits on the neutrino lifetimes, and assuming a hierarchical neutrino mass scheme where $m_1 \gg m_3$ and $m_2 \gg m_3$.  The translation of lower limits on the lifetime into upper limits on the combined couplings is in the main text.  Presently, with $N_{\rm obs} = 1$, the 90\%~C.L.~upper limits on combined couplings are $4.77 \cdot 10^{-6} ({\rm eV}/m_1)$ for $\nu_1$ and $7.24 \cdot 10^{-6} ({\rm eV}/m_2)$ for $\nu_2$.  In the future, with $N_{\rm obs} = 2$, the limits will be $3.17 \cdot 10^{-7} ({\rm eV}/m_1)$ for $\nu_1$ and $4.41 \cdot 10^{-6} ({\rm eV}/m_2)$ for $\nu_2$.  For comparison, the limits coming from the decay of solar neutrinos\ \cite{Berryman:2014qha} are $4.07 \cdot 10^{-6} ({\rm eV}/m_1)$ for $\nu_1$ and $9.72 \cdot 10^{-6} ({\rm eV}/m_2)$ for $\nu_2$.  
 
\setcounter{figure}{0}
\renewcommand{\thefigure}{E\arabic{figure}}
\begin{figure*}[t!]
 \centering
 \includegraphics[width=0.495\columnwidth,height=0.495\columnwidth]{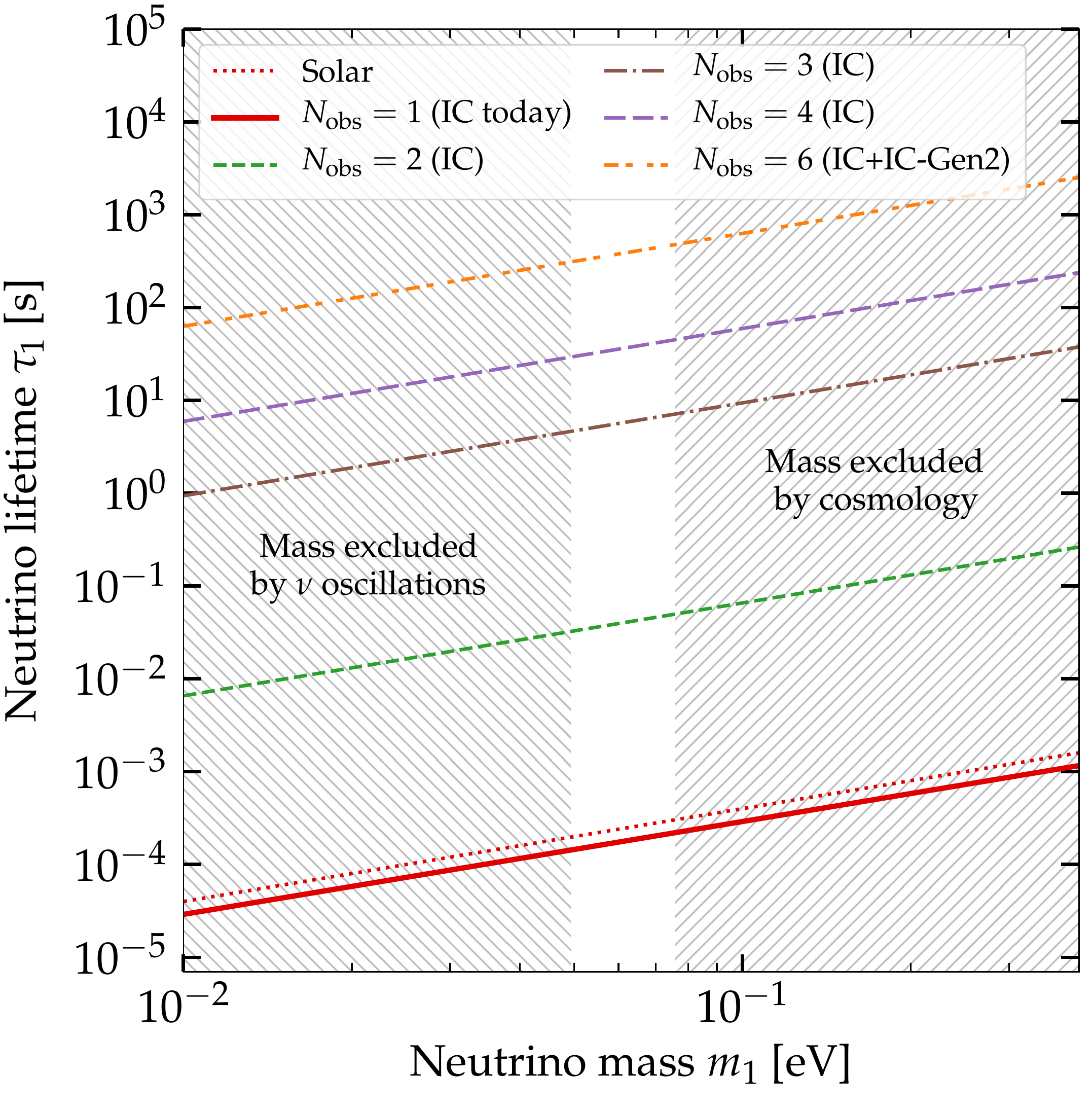}
 \includegraphics[width=0.495\columnwidth,height=0.495\columnwidth]{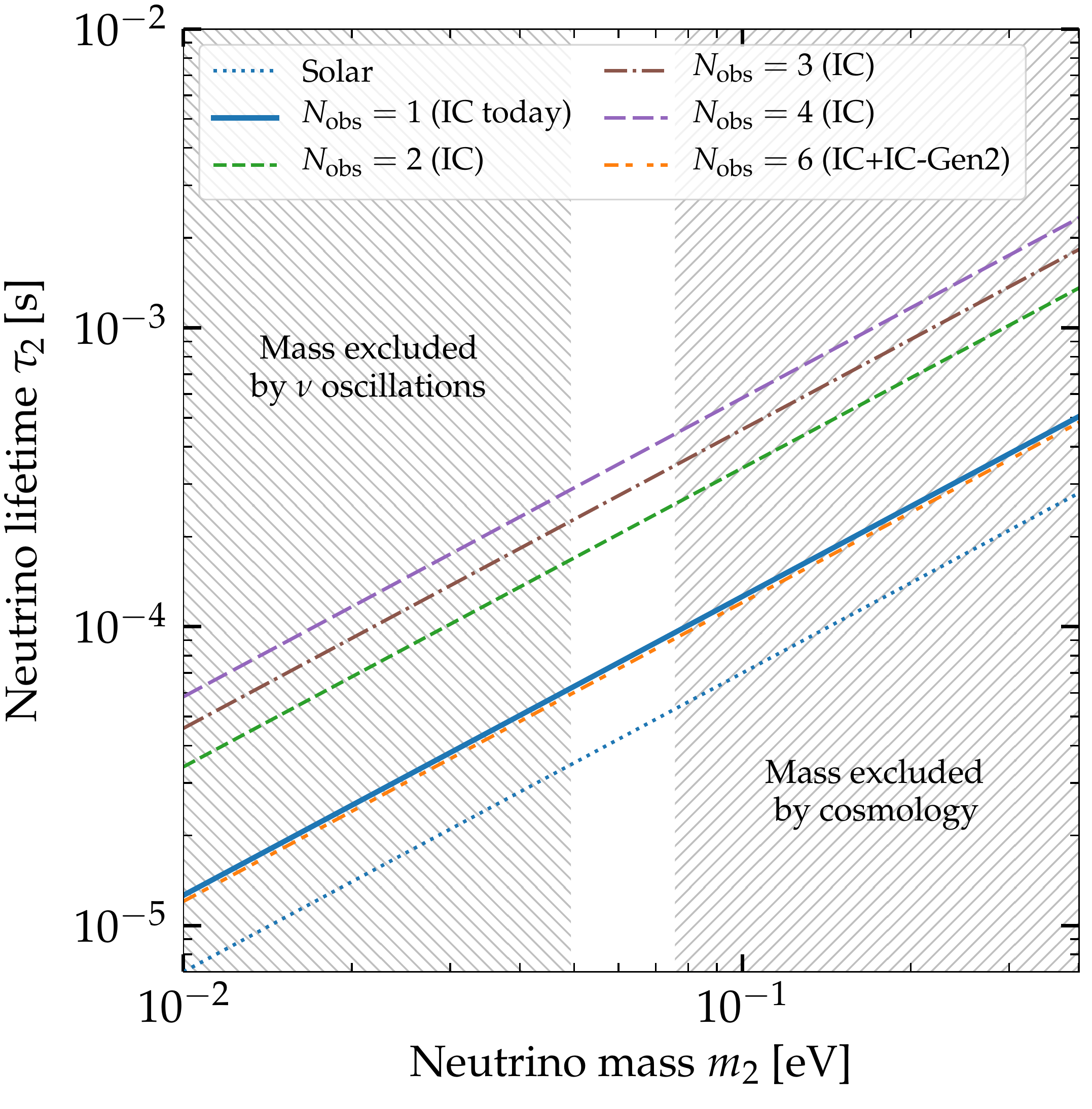}
 \caption{\label{fig:limits_future}Current and projected lower limits on the lifetimes of $\nu_1$ (left panel) and $\nu_2$ (right panel).  Note the different vertical scales between panels.  For the combined IceCube + IceCube-Gen2, we fix the mixing parameters at their current best-fit values.  Limits from solar neutrinos are shown for comparison.}
\end{figure*}

\renewcommand{\thefigure}{E\arabic{figure}}
\begin{figure}[t!]
 \centering
 \includegraphics[width=0.5\columnwidth,height=0.5\columnwidth]{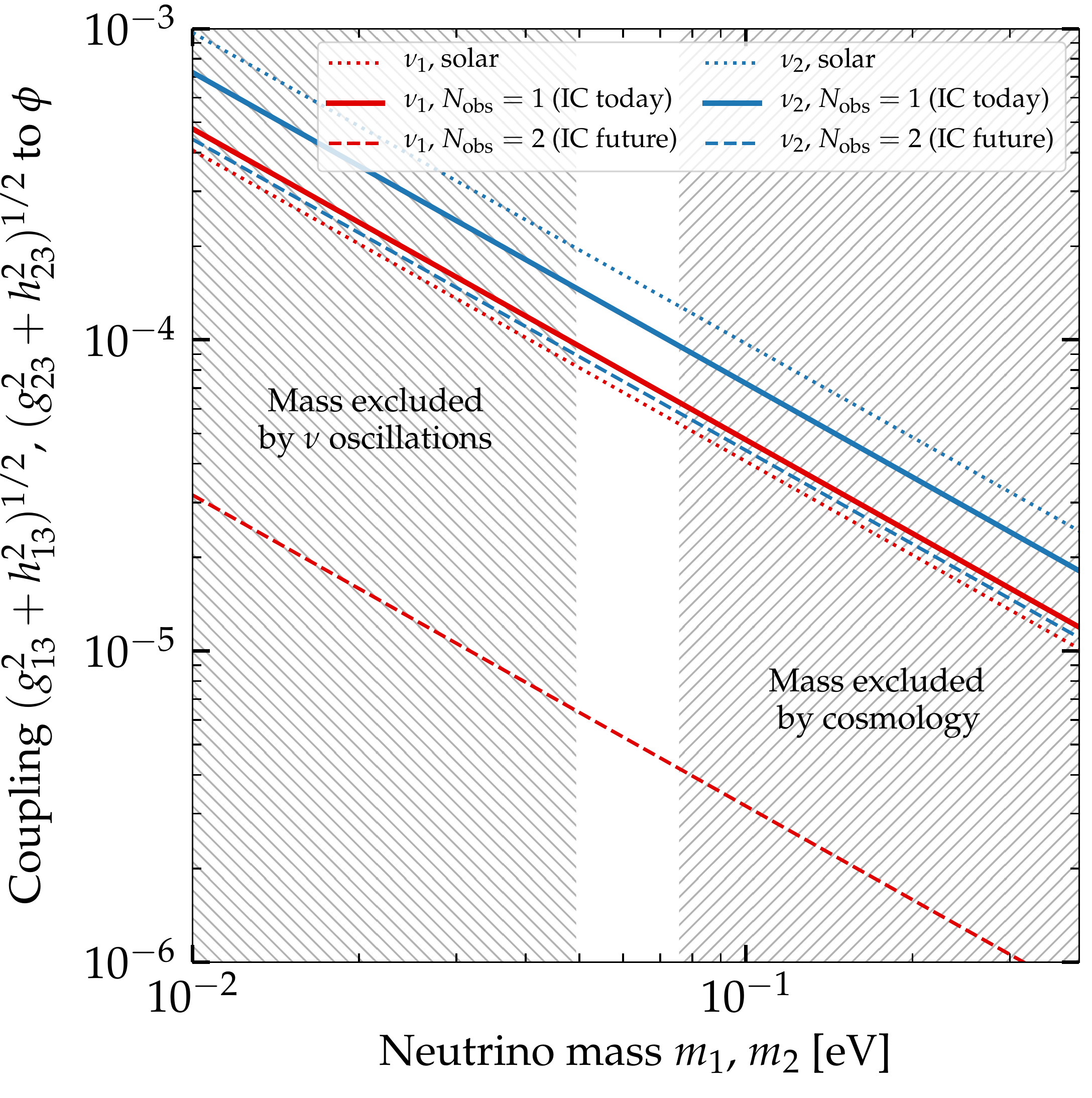}
 \caption{\label{fig:limits_g_m}
 Present-day and projected upper limits on the combined scalar ($g_{j3}$) and pseudoscalar ($h_{j3}$) couplings of $\nu_1$ and $\nu_2$ to a light new boson $\phi$.  Limits from solar neutrinos are shown for comparison.}
\end{figure}

\end{document}